\documentclass[useAMS,usenatbib,twocolumn]{mn2e}
\usepackage{graphicx}
\usepackage{subfigure}
\usepackage[fleqn]{amsmath}

\newcommand{\mpt}{\mathrm{.}}
\newcommand{\mcm}{\mathrm{,}}

\renewcommand{\Vec}[1]{ \mbox{\boldmath$ #1 $} }

\newcommand{\apjl}{ApJ}
\newcommand{\apj}{ApJ}

\newcommand{\mnras}{MNRAS}

\newcommand{\aj}{AJ}

\title[Lensing degeneracies and mass substructure]{Lensing degeneracies and mass substructure}
\author[J. Liesenborgs and S. De Rijcke]
{J. Liesenborgs$^1$\thanks{Corresponding author: jori.liesenborgs@uhasselt.be} and S. De Rijcke$^2$ \\
$^1$ Expertisecentrum voor Digitale Media, Universiteit Hasselt, Wetenschapspark 2, B-3590, Diepenbeek, Belgium \\ 
$^2$ Sterrenkundig Observatorium, Universiteit Gent, Krijgslaan 281, S9, B-9000, Gent, Belgium}

\begin{document}

\date{} 
	
	\pagerange{\pageref{firstpage}--\pageref{lastpage}} \pubyear{2012}
	
	\maketitle \label{firstpage} 

	\begin{abstract} 
		The inversion of gravitational lens systems is hindered by the
		fact that multiple mass distributions are often equally compatible
		with the observed properties of the images. Besides using clear examples
		to illustrate the effect of the so-called monopole and mass sheet 
		degeneracies, this article introduces the most general form 
		of said mass sheet degeneracy. While the well known version of this
		degeneracy rescales a single source plane, this generalization
		allows any number of sources to be rescaled. Furthermore, it shows how
		it is possible to rescale each of those sources with a different scale
		factor. Apart from illustrating that the mass sheet degeneracy is not broken
		by the presence of multiple sources at different redshifts, it will
		become apparent that the newly constructed mass distribution necessarily
		alters the existing mass density precisely at the locations of the images
		in the lens system, and that this change in mass density is linked to the
		factors with which the sources were rescaled. Combined with the fact that
		the monopole degeneracy introduces a large amount of uncertainty about the
		density in between the images, this means that both
		degeneracies are in fact closely related to substructure in the
		mass distribution. 
		An example simulated lensing situation based on an elliptical version of
		a Navarro-Frenk-White profile explicitly shows that such degeneracies are
		not easily broken by observational constraints, even when multiple sources
		are present.
		Instead, the fact that each lens inversion method
		makes certain assumptions, implicit or explicit, about the smoothness 
		of the mass distribution means that in practice the degeneracies are
		broken in an artificial manner rather than by observed properties of
		the lens system.
	\end{abstract}
	
	\begin{keywords}
		gravitational lensing -- dark matter
	\end{keywords}
	
	\section{Introduction}

		Not only can strong gravitational lensing yield impressive images, from
		multiply imaged quasars over deformed galaxies to partial or full
		Einstein rings, it is also an invaluable tool to estimate the mass
		of the deflector. Moreover, since the precise positions and deformations
		of the images of a source depend on the exact distribution of the mass,
		gravitational lensing even holds the promise of constraining the shape
		of said mass distribution. 
		
		In principle, this strong lens inversion sounds fairly straightforward: use
		a particular model and optimize its parameters so that it reproduces the
		observations as well as possible. In practice however, one is hindered
		by gravitational lensing degeneracies which allow a wide range of mass
		distributions to produce the exact same image configuration. On the
		level of parametric lens inversion results, this can manifest itself
		as parameter degeneracies (e.g. \citet{1994AJ....108.1156W}), but at their core the degeneracies
		are or course present at the level of the mass distribution. There,
		the mass sheet degeneracy (e.g. \citet{FalcoMassSheet} or
		\citet{SahaSteepness}) will rescale the projected mass distribution while
		adding a constant density mass sheet (or disc), only affecting the time
		delays between the images. Recently it was shown that a similar procedure
		is still possible when multiple sources at different redshifts are
		observed \citep{Liesenborgs3}. While this degeneracy therefore can still
		be broken when time delay measurements are present,
		the situation is far worse with respect to the so-called monopole
		degeneracy \citep{Liesenborgs4}. This type of degeneracy can be used
		to redistribute the mass in between the images, without changing any
		of the observable properties of the images.
	
		By presenting a further generalization of this mass sheet degeneracy,
		this article will illustrate an important relation between different degenerate
		versions of a lensing mass distribution. This will illustrate
		how degeneracies are closely related to the substructure in the
		mass density, which, in turn, offers insights into what one may
		constrain about the lensing mass. 

		After briefly introducing the lensing formalism and used notations in
		section \ref{sec:formalism}, the basics of the monopole and mass sheet
		degeneracies are reviewed in section \ref{sec:basics}. This will serve 
		as the starting point for some extensions in section \ref{sec:extensions}
		after which an experiment in section \ref{sec:experiment} will illustrate
		precisely how difficult it can be to estimate the parameters of a specific
		type of mass distribution, an elliptical generalization of a Navarro-Frenk-White
		profile \citep{1996ApJ...462..563N}.
		Finally, in section \ref{sec:conclusions} we shall discuss what the implications 
		of these degeneracies are and what different types of information can 
		help constrain.
	
	\section{Lensing formalism}\label{sec:formalism}

		In this section, the same notation as \citet{SchneiderBook} is used, which
		the interested reader can consult for a thorough review.
		The lens equation (\ref{eq:lenseq}) in essence states that due to the gravitational
		deflection $\Vec{\hat{\alpha}}$ of a light ray, when looking in a direction 
		$\Vec{\theta}$ one sees what would be observed in direction $\Vec{\beta}$ if
		the lens effect could be turned off:
		\begin{equation}\label{eq:lenseq}
			\Vec{\beta}(\Vec{\theta}) = \Vec{\theta} - 
						\frac{D_{\rm ds}}{D_{\rm s}}\Vec{\hat{\alpha}}(\Vec{\theta})\mpt
		\end{equation}
		The deflection angle $\Vec{\hat{\alpha}}$ depends on the two-dimensional projected
		mass distribution $\Sigma$ in the following fashion:
		\begin{equation}
			\Vec{\hat{\alpha}} = \frac{4 G D_{\rm d}}{c^2}
			     \int\frac{\Sigma(\Vec{\theta'})(\Vec{\theta}-\Vec{\theta'})}
			               {\left|\Vec{\theta}-\Vec{\theta'}\right|^2}
						   d\Vec{\theta'} \mpt
		\end{equation}
		In these equations, the geometry of the lensing scenario is given by the angular
		diameter distances $D_{\rm d}$, $D_{\rm s}$ and $D_{\rm ds}$, describing the
		distance to the lens, to the source and the distance between lens and source
		respectively. It will often be convenient to use the so-called critical density:
		\begin{equation}\label{eq:critdens}
			\Sigma_{\rm cr} = \frac{c^2}{4\pi G}\frac{D_{\textrm{s}}}{D_{\textrm{ds}} D_{\textrm{d}}} \mcm
		\end{equation}
		which depends on the redshift of a source through the angular diameter distances
		involved.

		In the special case of a circularly symmetric mass distribution, for convenience
		assumed to be centered on the origin of the coordinate system, the expression of
		the deflection angle reduces to:
		\begin{equation}\label{eq:alphasymm}
			\Vec{\hat{\alpha}}(\Vec{\theta}) = \frac{4 G M(\theta)}
			{c^2 D_{\rm d} \theta^2}\Vec{\theta}\mcm
		\end{equation}
		in which $M(\theta)$ represents the mass enclosed within radius $\theta$:
		\begin{equation}\label{eq:enclosedmass}
			M(\theta) = 2\pi D_{\rm d}^2\int_0^\theta\Sigma(\theta')\theta' d\theta'\mpt
		\end{equation}

		The deflection angle can be shown to be proportional to the gradient of the 
		so-called lens potential
		\begin{equation}\label{eq:lenspotential}
			\Vec{\nabla}\psi(\Vec{\theta}) = \frac{D_{\rm ds}}{D_{\rm s}}\Vec{\hat{\alpha}}(\Vec{\theta}) \mcm
		\end{equation}
		which is related to the time light takes to travel from the source at
		position $\Vec{\beta}$ to an image at position $\Vec{\theta}$:
		\begin{equation}\label{eq:timedelay}
			t\left(\Vec{\theta}\right) = \frac{1+z_{\rm d}}{c}\frac{D_{\rm d}D_{\rm s}}{D_{\rm ds}}
			\left(\frac{1}{2}\left(\Vec{\theta}-\Vec{\beta}\right)^2-\psi(\Vec{\theta})\right) + \mathrm{const}\mpt
		\end{equation}
		In this last equation, $z_{\rm d}$ is the redshift of the gravitational
		lens. Since, in practice one is only interested in the difference in
		time delay between images of the same source, the constant in this
		expression is of no importance. The curvature of the lens potential
		can be shown to represent the mass density at the corresponding point:
		\begin{equation}\label{eq:convergence}
			\frac{1}{2}\nabla^2\psi\left(\Vec{\theta}\right) = \frac{\Sigma\left(\Vec{\theta}\right)}{\Sigma_{\rm cr}} \mpt
		\end{equation}

		Finally, the magnification $\mu$ that an image at position $\Vec{\theta}$ experiences,
		can be calculated from the lens equation as follows:
		\begin{equation}\label{eq:magnification}
			\mu^{-1}\left(\Vec{\theta}\right) = \left|\frac{\partial \beta_i}{\partial \theta_j}\right|\mcm
		\end{equation}
		i.e. it is the jacobian determinant of this equation.

	\section{Degeneracy basics}\label{sec:basics}
		
		\subsection{Monopole degeneracy}

			\begin{figure*}
				\centering
				\subfigure{\includegraphics[width=0.49\textwidth]{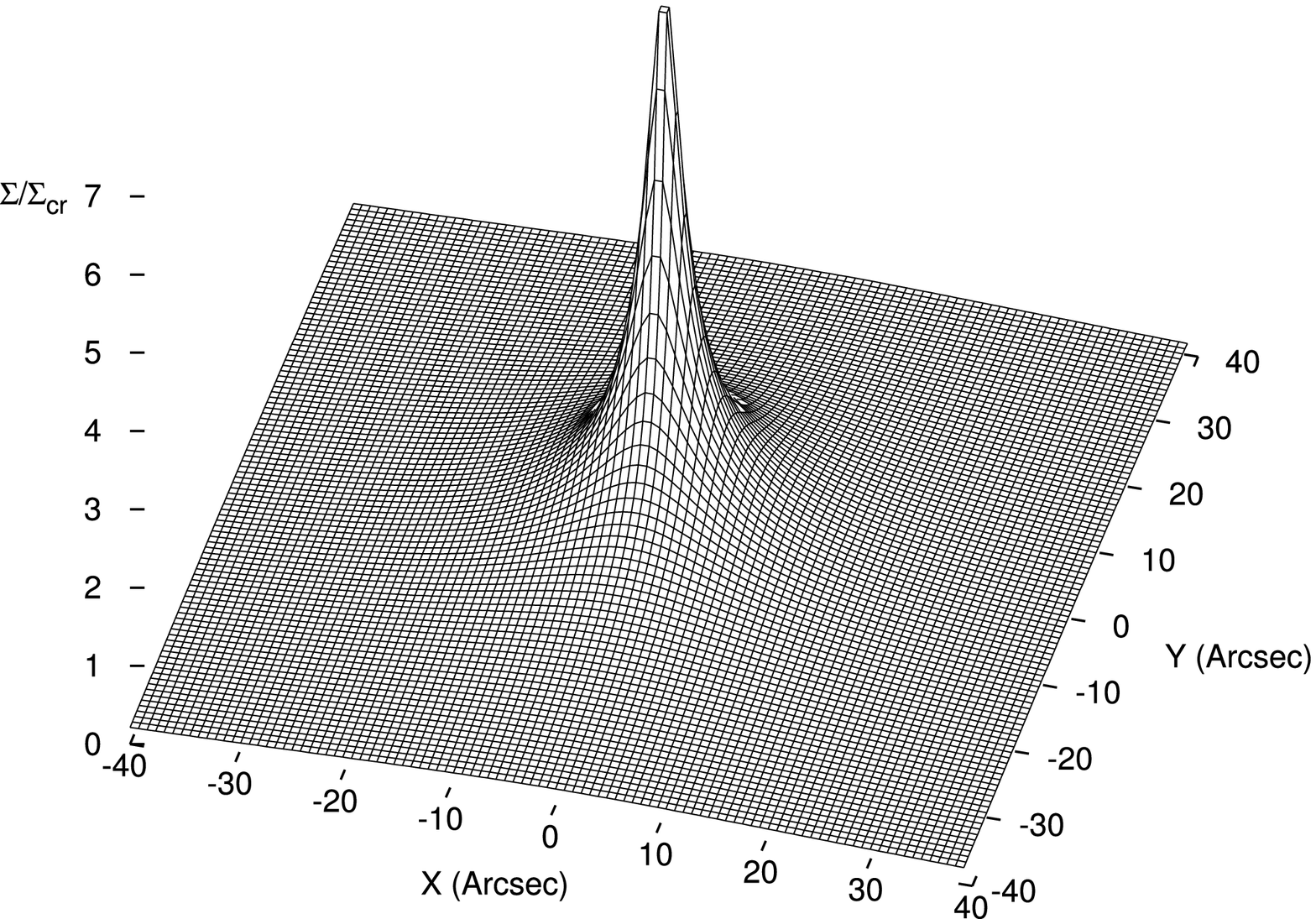}}
				\subfigure{\includegraphics[width=0.49\textwidth]{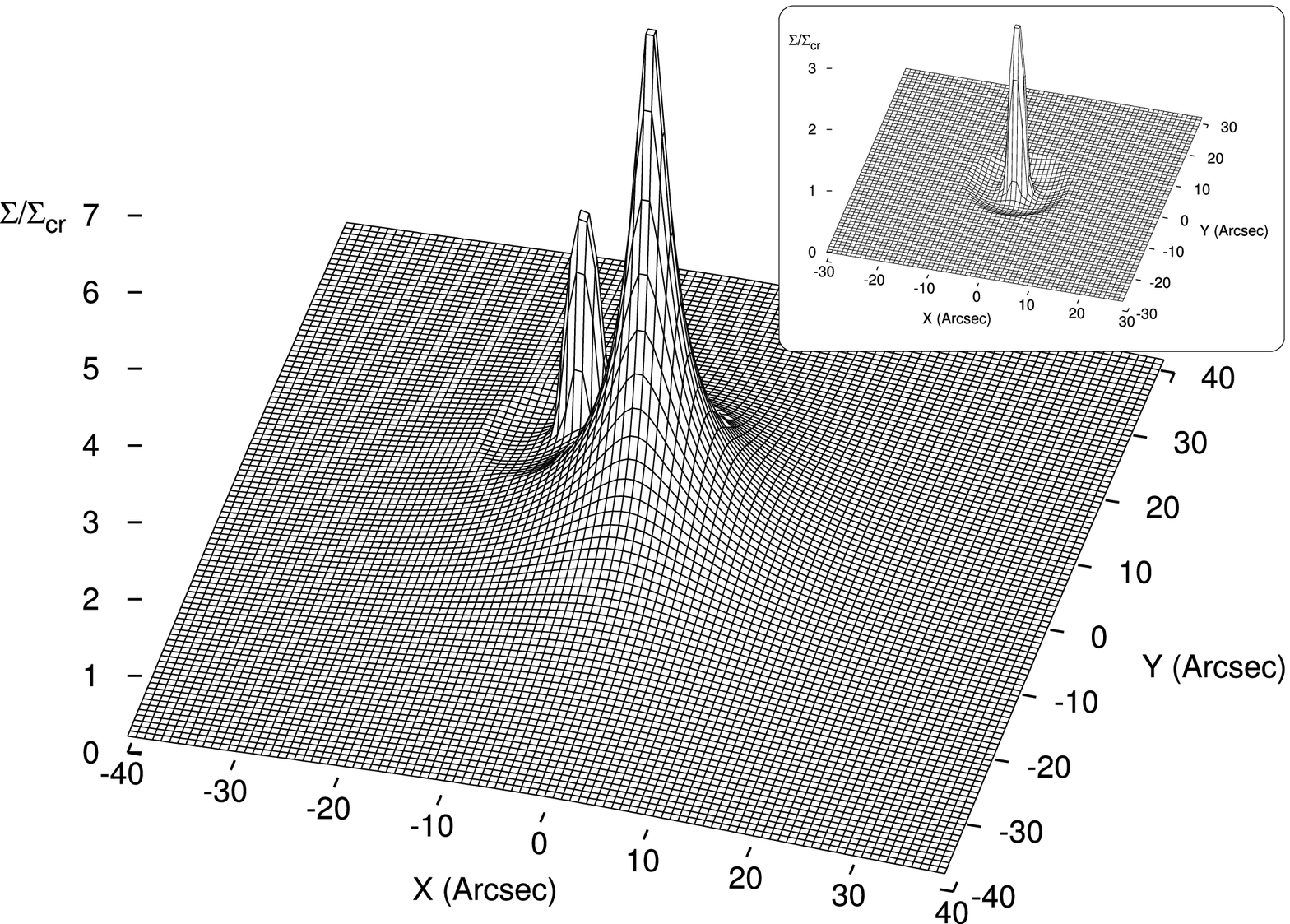}}
				\caption{Left panel:~this non-singular isothermal ellipse based
						 gravitational lens transforms a single source into the five images that can be 
						 seen in the left panel of 
						 Fig.~\ref{fig:monopoleeffect}. Right panel:~when the mass distribution shown in the
						 inset is added in the region enclosed by a circle in Fig.~\ref{fig:monopoleeffect} 
						 (left panel), this mass distribution is created. Note that although the added mass
						 distribution itself contains a region of negative density, the total density of the
						 lens is positive everywhere.}
				\label{fig:monopole}
			\end{figure*}
			\begin{figure*}
				\centering
				\subfigure{\includegraphics[width=0.49\textwidth]{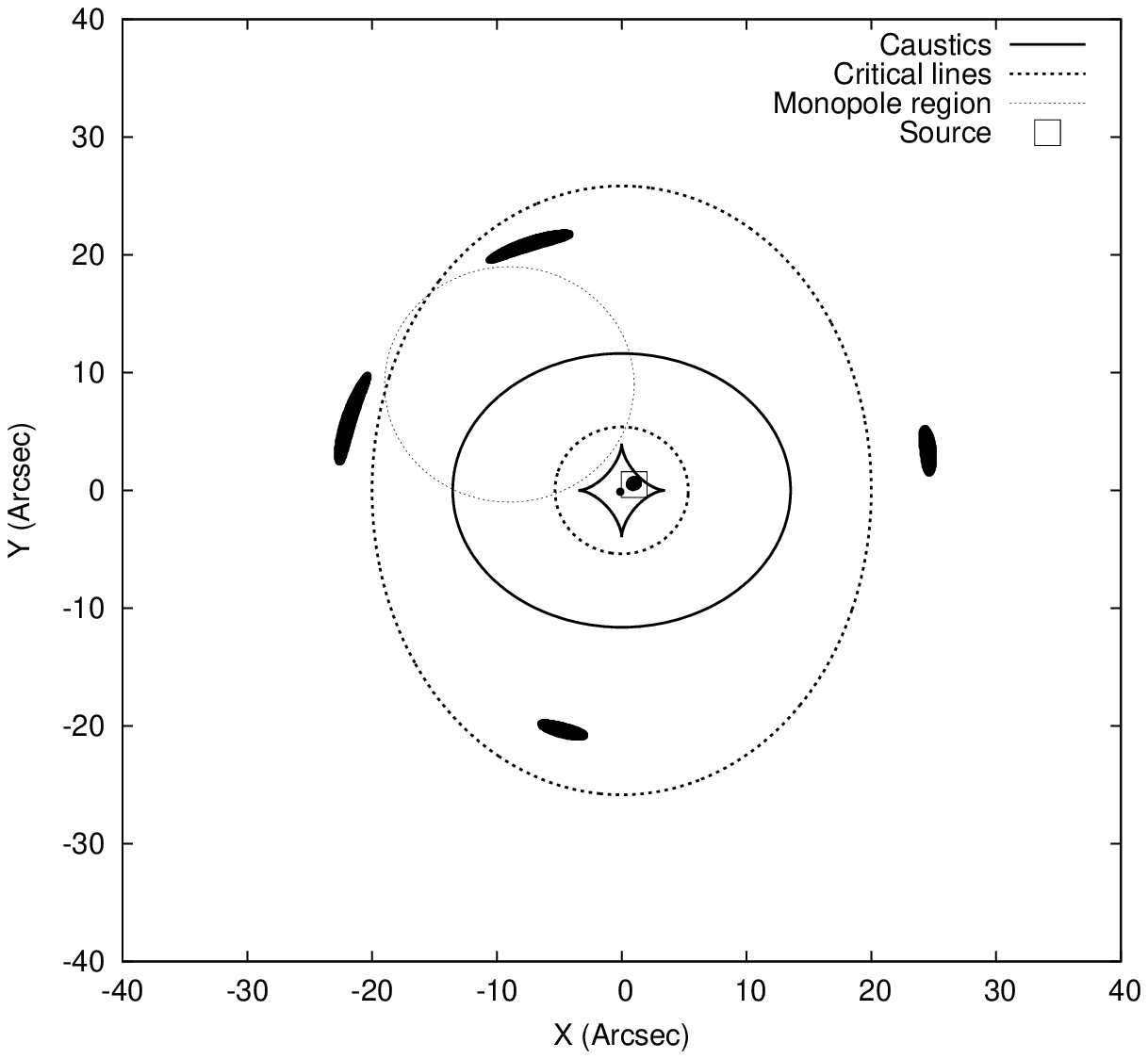}}
				\subfigure{\includegraphics[width=0.49\textwidth]{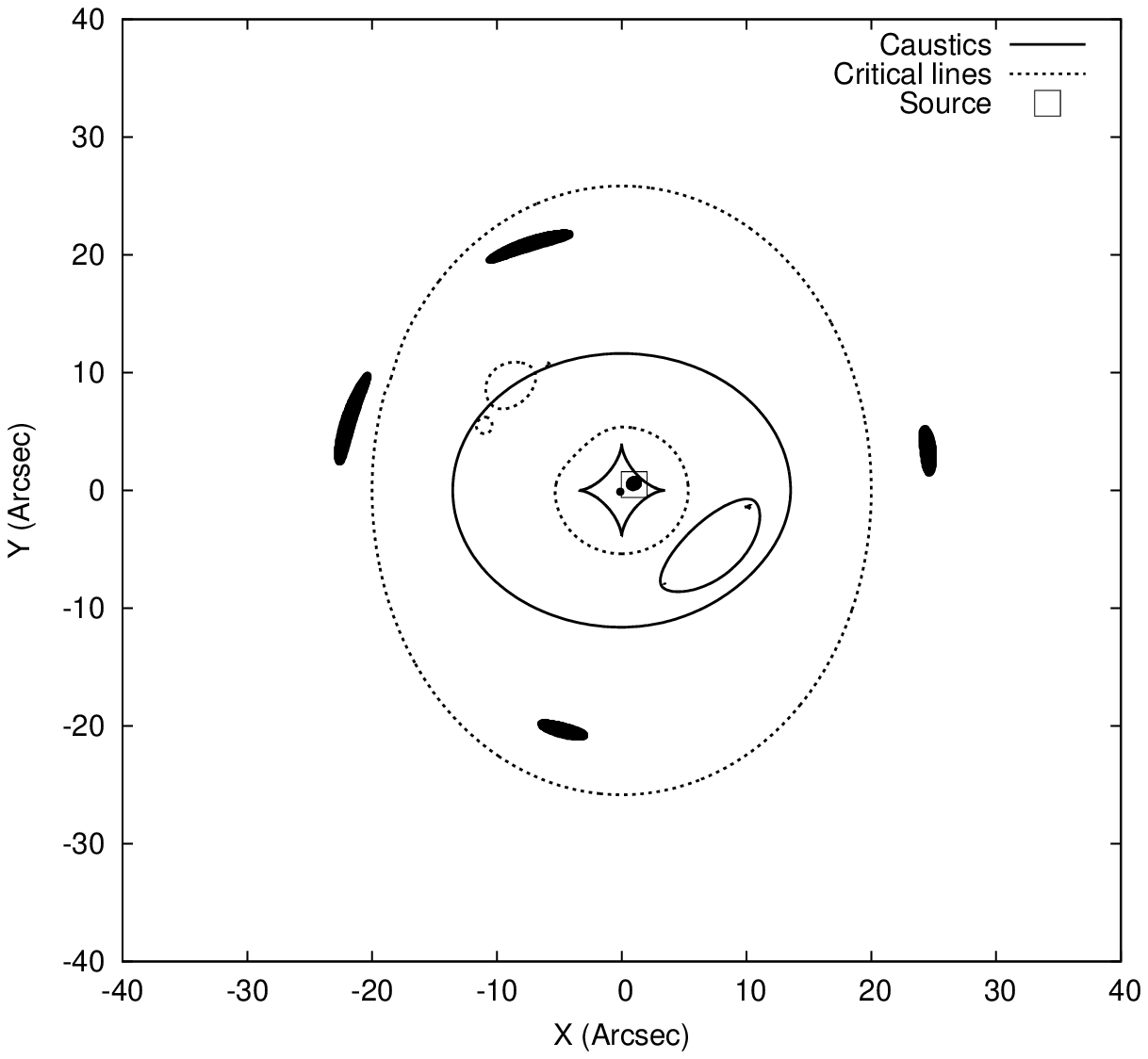}}
				\caption{Left panel:~the gravitational lens shown in the left panel of Fig.~\ref{fig:monopole}
				         creates five images of the source enclosed by a small square. A circle encloses a 
						 region which does not contain any of the images, and can therefore be used to
						 illustrate the monopole degeneracy (see text). Right panel:~when the circularly
						 symmetric mass distribution from the inset of Fig.~\ref{fig:monopole} is 
						 added in the circular region, the resulting mass distribution
						 (right panel of Fig.~\ref{fig:monopole}) still maps the source onto the exact same
						 images. Note that the change in the critical lines and caustics illustrates the
						 fact that inside the circular region, the lens equation indeed has changed.}
				\label{fig:monopoleeffect}
			\end{figure*}

			The so-called monopole degeneracy, first mentioned in \citet{Saha2000} and studied
			in more detail in \citet{Liesenborgs4}, is 
			both the least complex and most elusive degeneracy one can think of. 
			Consider the mass distribution shown in the inset in Fig.~\ref{fig:monopole}. 
			Being a circularly symmetric mass distribution, its lens effect is governed by
			equation~(\ref{eq:alphasymm}). Since this particular mass distribution was
			constructed to have zero enclosed mass outside a particular radius, the same
			equation shows that it will not cause a deflection angle outside said radius.
			This kind of mass density will be referred to as a monopole, since it can be
			described by only the monopole term in a multipole expansion.
	
			An example will illustrate the importance of such a monopole. The elliptical
			mass distribution shown in the left panel of Fig.~\ref{fig:monopole}
			transforms a single source into five distinct images, as can be seen in the
			left panel of Fig.~\ref{fig:monopoleeffect}. Indicated in the same figure, is
			a circular region in which none of the images are located, and which has the
			precise same diameter as the aforementioned monopole.
			When adding the monopole in this location, this implies that the
			lens equation will not be affected outside the circular region; only within
			will the deflection of light be affected. That this is indeed the case, can
			clearly be seen in the right panel of Fig.~\ref{fig:monopoleeffect}. There,
			the lens effect is calculated for the same source, but for the combined mass distribution
			shown in the right panel of Fig.~\ref{fig:monopole}. As was expected, the exact same
			five images are generated by this clearly different mass density. The fact
			that the lens equation was modified inside the circular region can be seen
			in the change in critical lines in this region. Since the deflection is different
			inside the circular area, in principle it is even possible that extra images are
			generated there. However, this particular example shows that this does 
			not necessarily need to be the case, even for a rather large modification of
			the mass distribution.
	
			This example already shows that by `borrowing' some mass from a relatively
			large region, it is easy to introduce (or remove) a mass peak in a mass
			density, without changing the images that are generated. As was explicitly shown in
			\citep{Liesenborgs4} however, it is also possible to use such monopoles
			as basis functions which allows a complex redistribution of the mass by
			adjusting their weights. As long as no basis function overlaps with an image,
			none of the images will be affected by this operation. Some care needs
			to be taken not to introduce additional, unobserved images though.
	
			From the discussion above, it is already trivially clear that this type
			of degeneracy does not modify the image positions generated by a
			particular source. The fact that the deflection angle is zero outside
			a certain radius, by equation~(\ref{eq:magnification}) implies that the
			magnification at the location of the images will not be affected either.
			Equation~(\ref{eq:lenspotential}) on the other hand shows that the lens
			potential outside said radius may change by a constant value. However,
			when measuring time delays between images, such a constant offset in the
			potential will have	no effect, as can be seen in equation~(\ref{eq:timedelay}).

		\subsection{The mass sheet degeneracy}

			The mass-sheet degeneracy is undoubtedly
			one of the most famous degeneracies in gravitational lensing. Contrary
			to what the name may suggest, to obtain a degenerate solution it is not
			sufficient to simply add a sheet of mass to an existing mass distribution.
			Instead, one has to rescale the original mass distribution as well, prompting
			the use of the alternative name of steepness degeneracy \citep{SahaSteepness}.

			\begin{figure}
				\centering
				\includegraphics[width=0.48\textwidth]{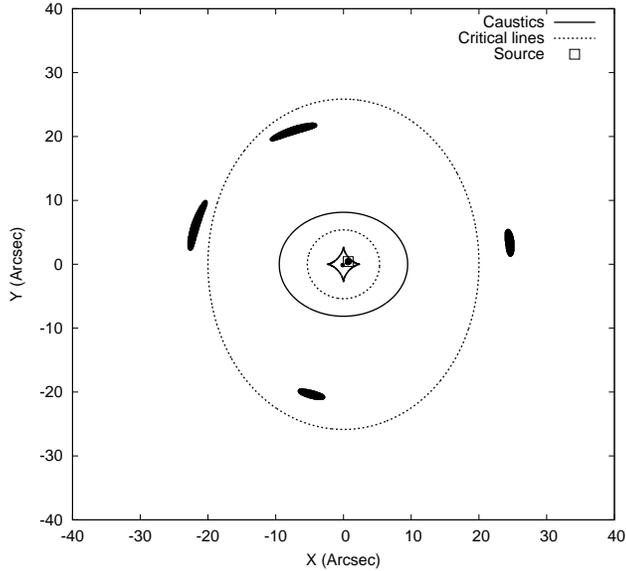}
				\caption{The mass sheet degeneracy in its simplest form causes a rescaled
						 version of the source plane to correspond to the same images. The fact that the
						 entire source plane has been rescaled is most obvious when comparing the
						 caustics to those of the left panel of Fig. \ref{fig:monopoleeffect}. 
						 In this particular example, a value of $\lambda = 0.7$ was used in the construction 
						 of a degenerate solution according to equation (\ref{eq:sheetdegen}).}
				\label{fig:sheetdegenimages}
			\end{figure}

			As a concrete example, let us call the mass distribution of the left panel
			of Fig.~\ref{fig:monopole} $\Sigma_0$, producing a deflection field $\Vec{\hat{\alpha}}_0$.
			Using equation~(\ref{eq:alphasymm}), one can show that for a sheet of mass
			of precisely the critical density $\Sigma_{\rm cr}$, the deflection angle simply becomes:
			\begin{equation}\label{eq:sheetdeflection}
				\Vec{\hat{\alpha}}_{\rm s}\left(\Vec{\theta}\right) = \frac{D_{\rm s}}{D_{\rm ds}}\Vec{\theta}\mpt
			\end{equation}
			If one then constructs a new mass density $\Sigma_1$ as follows:
			\begin{equation}\label{eq:sheetdegen}
				\Sigma_1\left(\Vec{\theta}\right) = \lambda \Sigma_0\left(\Vec{\theta}\right) + (1-\lambda) \Sigma_{\rm cr} \mcm
			\end{equation}
			it is a straightforward exercise to show that for the combined deflection
			$\lambda\Vec{\hat{\alpha}}_0+(1-\lambda)\Vec{\hat{\alpha}}_{\rm s}$ the
			new lens equation is:
			\begin{equation}\label{eq:sheetdegenlenseq}
				\Vec{\beta}_1\left(\Vec{\theta}\right) = \lambda \Vec{\beta}_0\left(\Vec{\theta}\right)\mcm
			\end{equation}
			i.e. merely a rescaling of the original one. This means that a rescaled version
			of the source plane corresponds to the same images, as is illustrated in
			Fig.~\ref{fig:sheetdegenimages}.

			If the source involved is variable, it may be possible to measure
			time delays between the images:
			\begin{equation}
				\Delta t_{ij} = t(\Vec{\theta}_i) - t(\Vec{\theta}_j) \mcm
			\end{equation}
			where $\Vec{\theta}_i$ and $\Vec{\theta}_j$ are the positions of two
			of the images. For a sheet of mass consisting of precisely the 
			critical density, the projected potential is 
			\begin{equation}
				\psi_s(\theta) = \frac{1}{2}\theta^2 \mcm
			\end{equation}
			so that the mass sheet degeneracy transforms an initial lens potential
			$\psi_0$ into
			\begin{equation}
				\psi_1(\Vec{\theta}) = \lambda\psi_0(\Vec{\theta}) +\frac{1-\lambda}{2}\theta^2 \mpt
			\end{equation}
			Noting that the mass sheet degeneracy changes a source position $\Vec{\beta}_0$
			to $\Vec{\beta}_1 = \lambda \Vec{\beta}_0$, a brief calculation shows
			that the relationship between the original time delay $\Delta t_{ij,0}$
			and the time delay of the degenerate version $\Delta t_{ij,1}$ 
			becomes:
			\begin{equation}\label{eq:sheetdelay}
				\Delta t_{ij,1} = \lambda \Delta t_{ij,0} \mpt
			\end{equation}
			This means that time delay measurements break this mass sheet degeneracy,
			since a particular version of the degeneracy corresponds to a particular
			time delay.

		\begin{figure}
			\centering
			\includegraphics[width=0.49\textwidth]{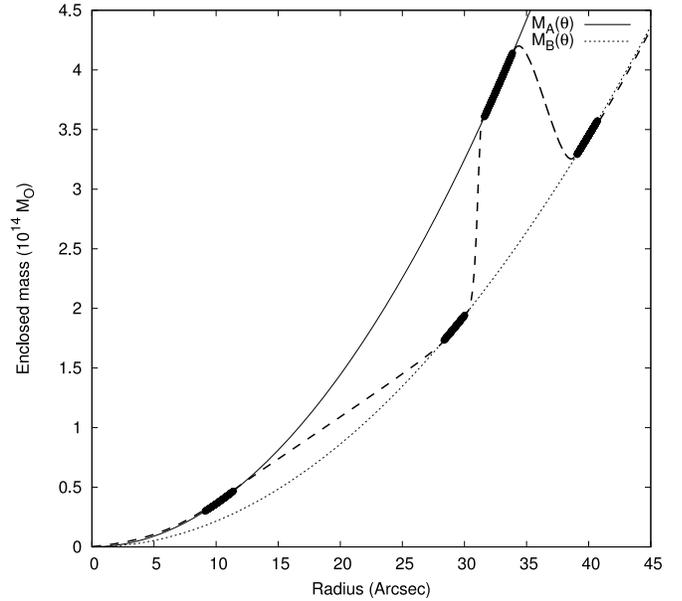}
			\caption{In this example, two sources at a different redshift
			         each produce two images due to the presence of a gravitational
					 lens. Because of the different redshifts, each source
					 would require a different critical density in the
					 construction of a solution that differs by the mass sheet 
					 degeneracy. The enclosed mass profiles corresponding
					 to these two densities are shown as curves $M_{\rm A}$
					 and $M_{\rm B}$. The locations of the images of each
					 source are indicated on these profiles. The dashed line, 
					 an interpolation between the two curves, shows a total mass 
					 profile that effectively behaves as the two sheets, from
					 the perspective of the images (see text).}
			\label{fig:doublesheet}
		\end{figure}

			The mass sheet degeneracy also has a particularly simple effect on the
			magnification factor $\mu$. Since each dimension is scaled by a factor
			$\lambda$, the surface area of a small source is scaled by a factor $\lambda^2$.
			Keeping the sizes of the images constant, this means that the new
			magnifications of the source are given by $\mu_1 = \lambda^{-2} \mu_0$.
			This result can also be obtained by combining the general expression for 
			the magnification (\ref{eq:magnification}) and the relation between the
			degenerate versions of the lens equation (\ref{eq:sheetdegenlenseq}).

	\section{Generalized mass sheet degeneracy}\label{sec:extensions}

		A first generalization of the mass sheet degeneracy was described in
		\citet{Liesenborgs3}. A brief review shall be presented here, before 
		discussing an even more general version of this degeneracy. It has
		often been claimed that having two distinct sources at different
		redshifts breaks the mass sheet degeneracy. From the construction
		above, the reason for this is clear. A sheet with precisely the
		critical density was used, but since the critical density is redshift 
		dependent, each source would 
		now require the	use of a sheet with a different density.

		This situation is illustrated in Fig.~\ref{fig:doublesheet}. Source $A$
		would require a sheet of mass which gives rise to the enclosed mass
		profile $M_{\rm A}(\theta)$. Similarly, source $B$ would need a sheet
		with mass profile $M_{\rm B}(\theta)$. The radii at which images of each 
		source are present are indicated on the corresponding mass profile.
		While it is true that a single mass sheet can no longer be used for
		a construction like the one in equation (\ref{eq:sheetdegen}), it indeed
		is possible to construct a mass density that for each source has the
		same effect as its corresponding mass sheet.

		The key point is that for a circularly symmetric mass distribution,
		equation~(\ref{eq:alphasymm}) shows that the deflection angle is caused 
		by the total enclosed mass within a specific radius, without any dependence
		on the structure of the mass profile within said radius. This implies
		that if one were to construct the mass profile indicated by the dashed
		line in Fig.~\ref{fig:doublesheet}, the deflection angle at the location
		of the images of source $A$ will be the same as when caused by a mass
		sheet of precisely its critical density. Similarly, the images of source $B$
		will effectively `see' a mass sheet with their critical density. The
		mass density corresponding to this new profile can then be used to
		construct a degenerate solution in the same way as in equation~(\ref{eq:sheetdegen}).
		Calling $R_{\rm A}$ and $R_{\rm B}$ the radii at which images of sources $A$ and
		$B$ are located, by construction this implies that:
		\begin{eqnarray}
			\Sigma_1\left(R_{\rm A}\right) & = & \lambda \Sigma_0\left(R_{\rm A}\right) + (1-\lambda) \Sigma_{\rm cr}\left(z_{\rm A}\right) \nonumber \\
			\Sigma_1\left(R_{\rm B}\right) & = & \lambda \Sigma_0\left(R_{\rm B}\right) + (1-\lambda) \Sigma_{\rm cr}\left(z_{\rm B}\right) \label{eq:doublesheetdegen} \mpt 
		\end{eqnarray}
		
		\begin{figure*}
			\centering
			\subfigure{\includegraphics[angle=-90,width=0.30\textwidth]{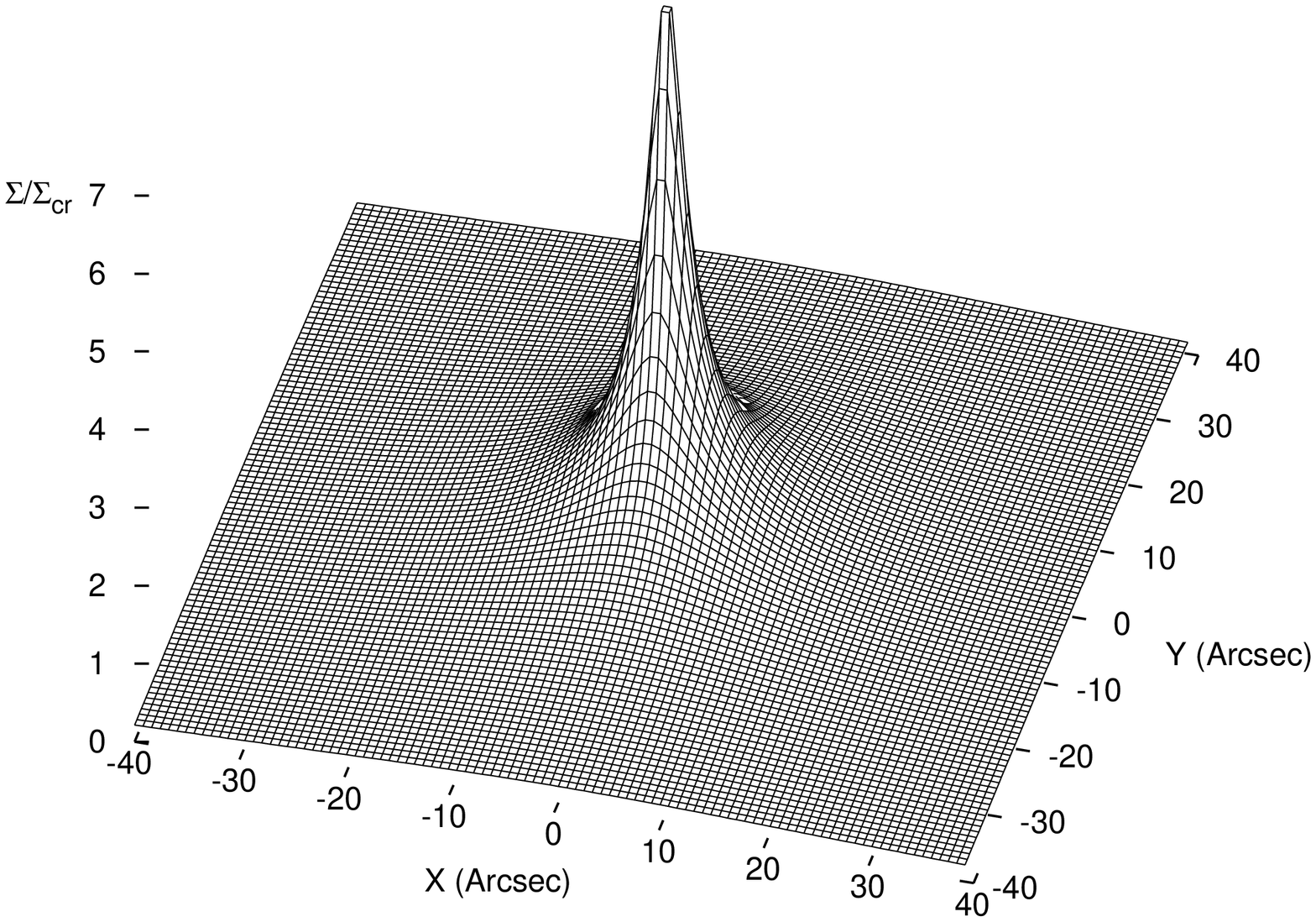}}
			\qquad
			\subfigure{\includegraphics[angle=-90,width=0.30\textwidth]{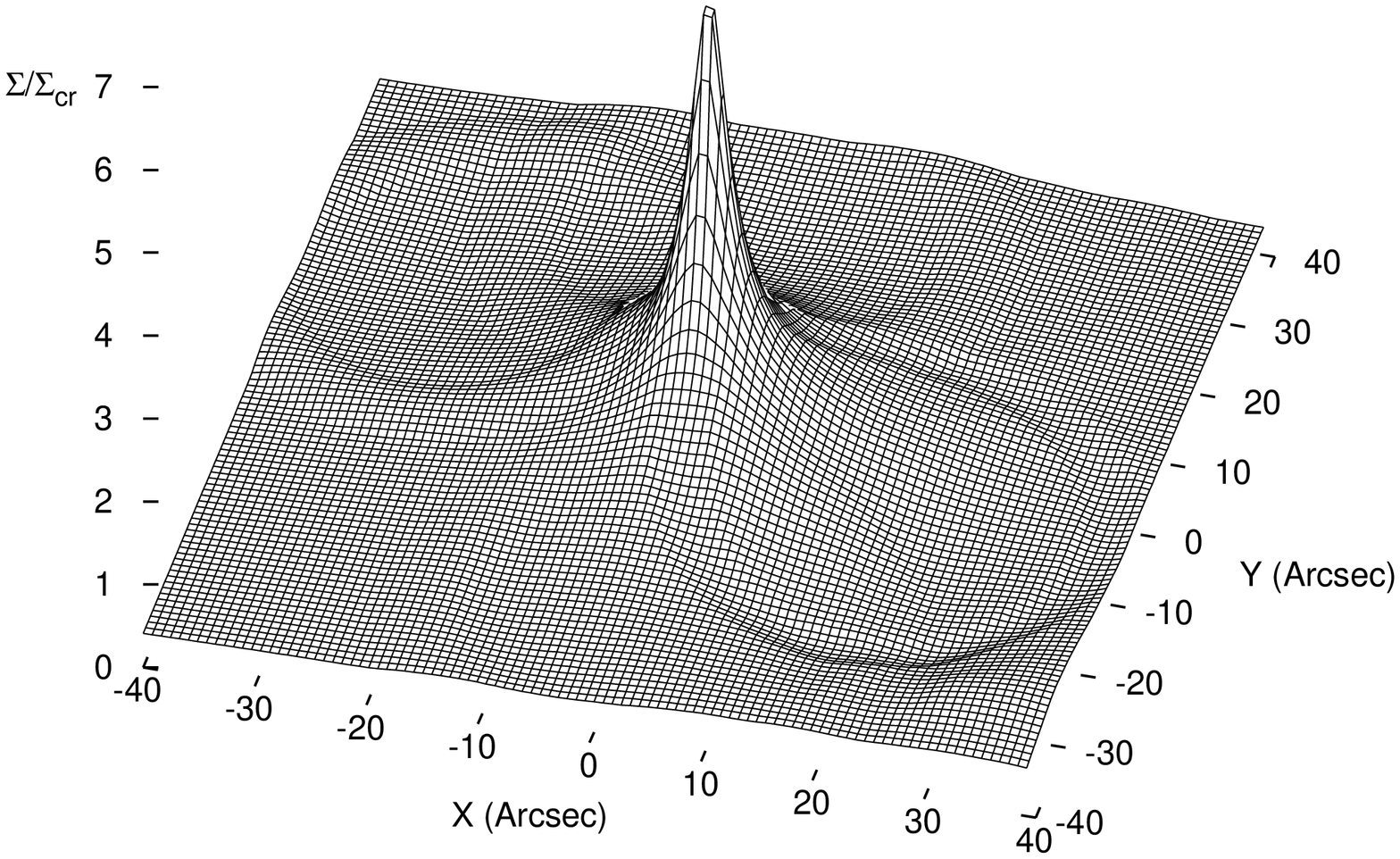}}
			\qquad
			\subfigure{\includegraphics[angle=-90,width=0.30\textwidth]{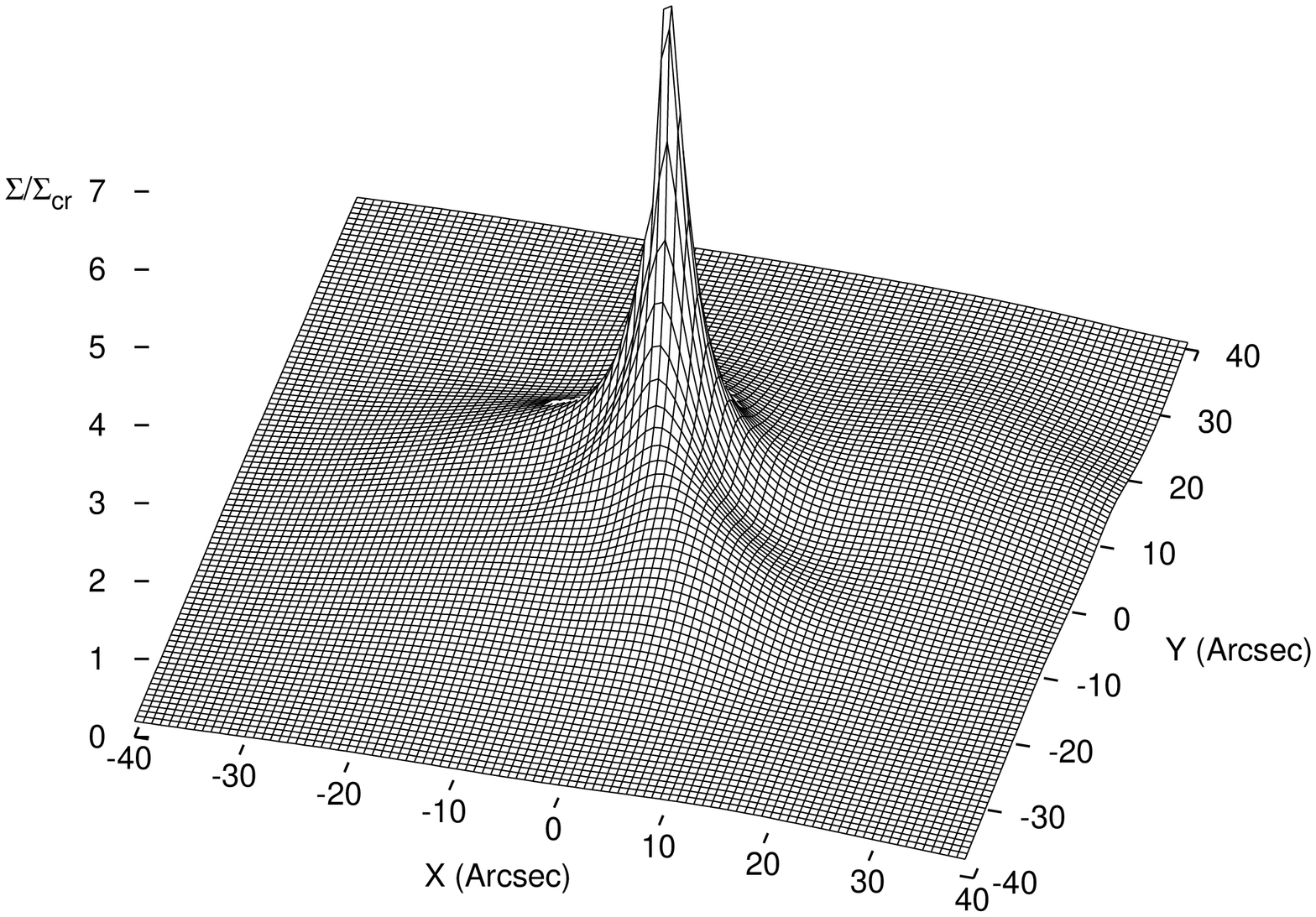}}
			\subfigure{\includegraphics[angle=0,width=0.30\textwidth]{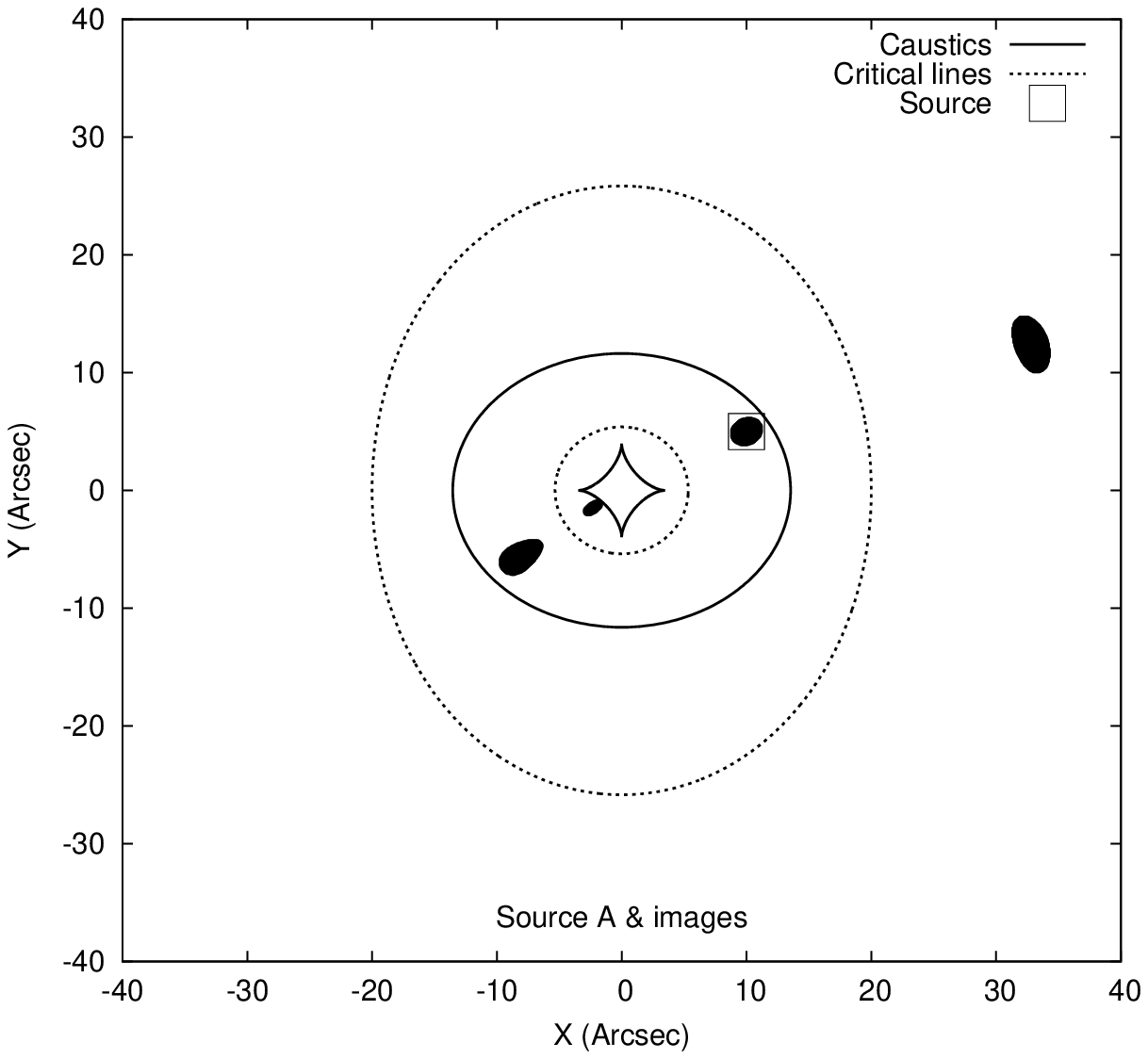}}
			\qquad
			\subfigure{\includegraphics[angle=0,width=0.30\textwidth]{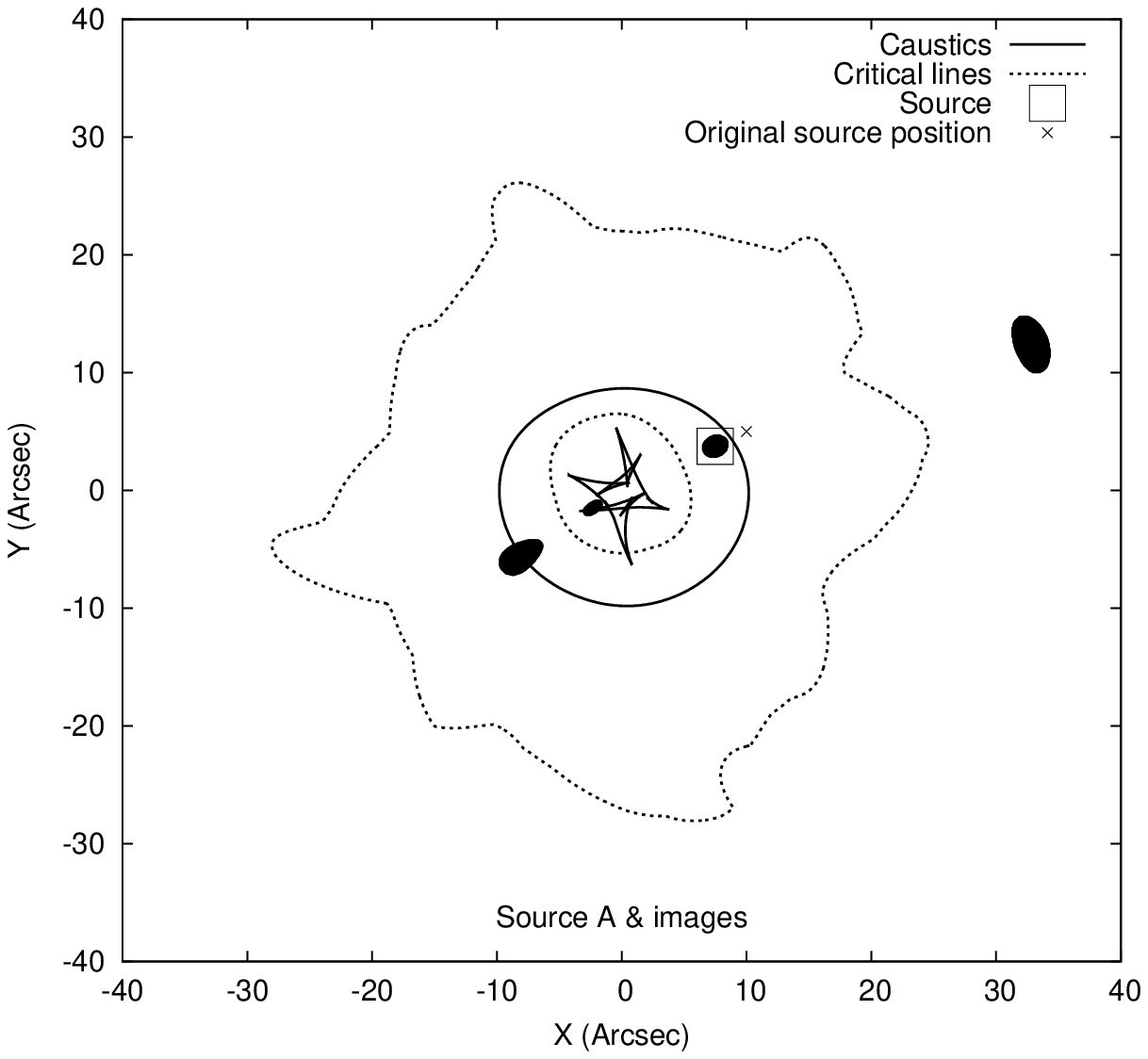}}
			\qquad
			\subfigure{\includegraphics[angle=0,width=0.30\textwidth]{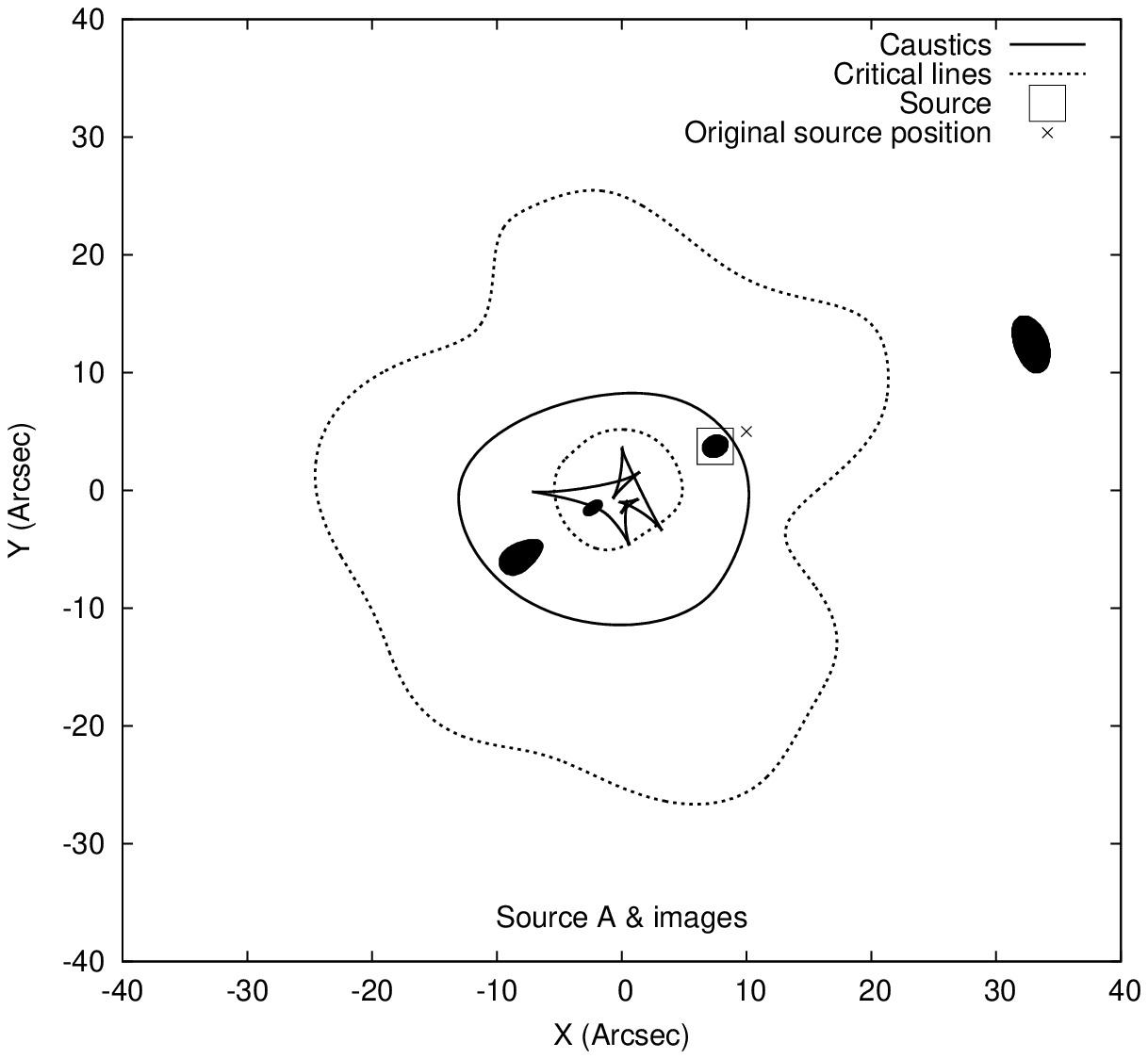}}
			\subfigure{\includegraphics[angle=0,width=0.30\textwidth]{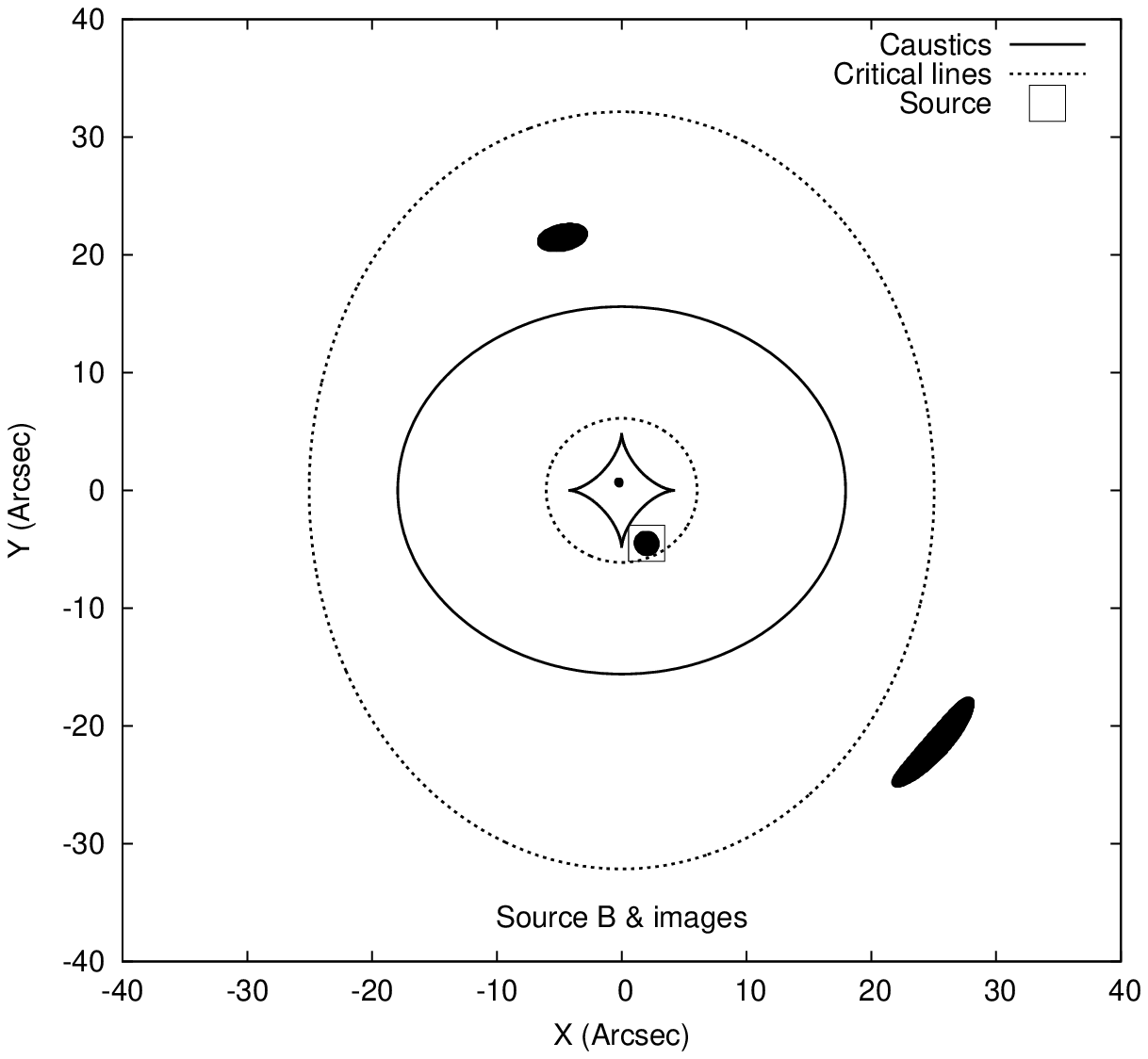}}
			\qquad
			\subfigure{\includegraphics[angle=0,width=0.30\textwidth]{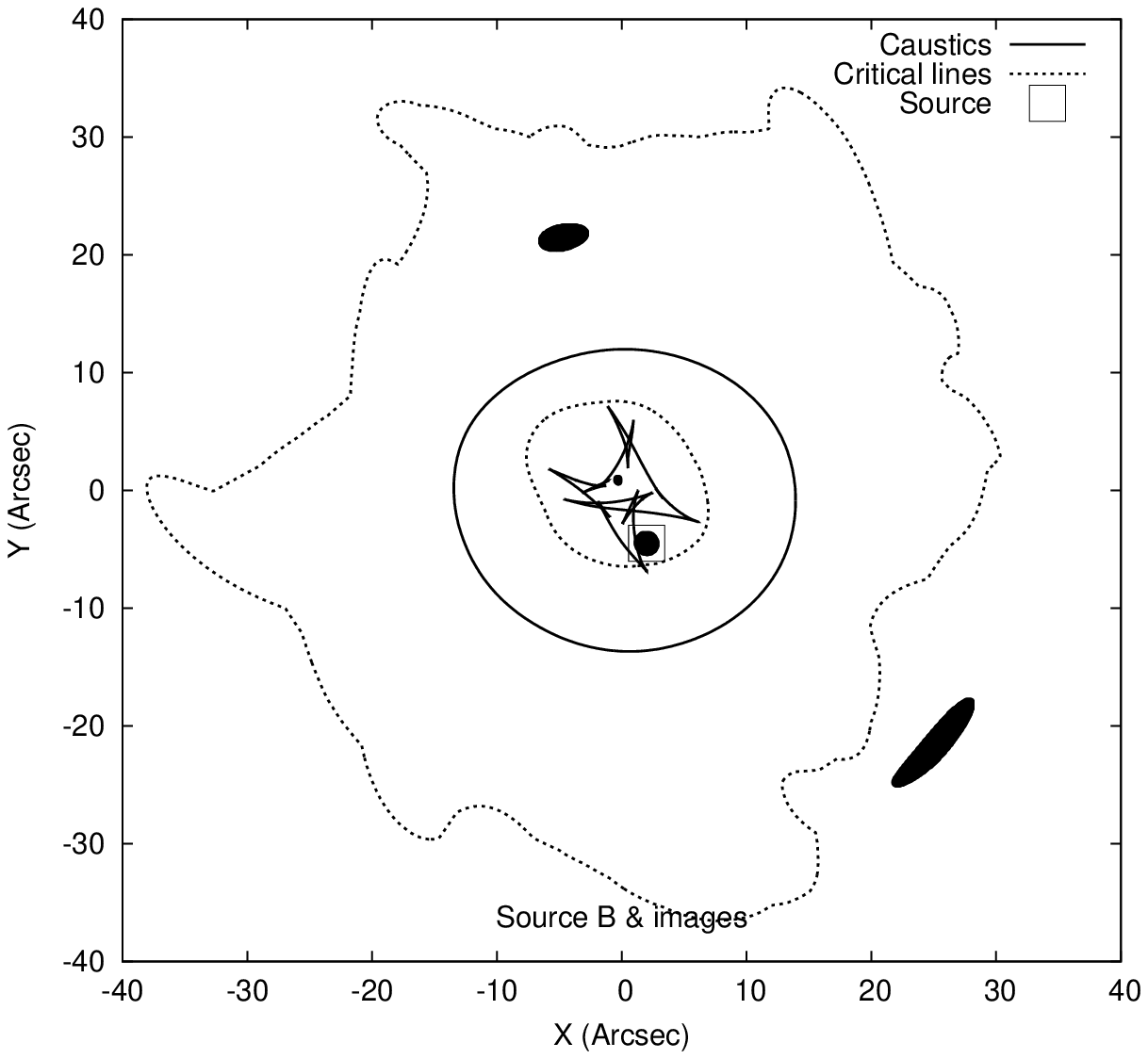}}
			\qquad
			\subfigure{\includegraphics[angle=0,width=0.30\textwidth]{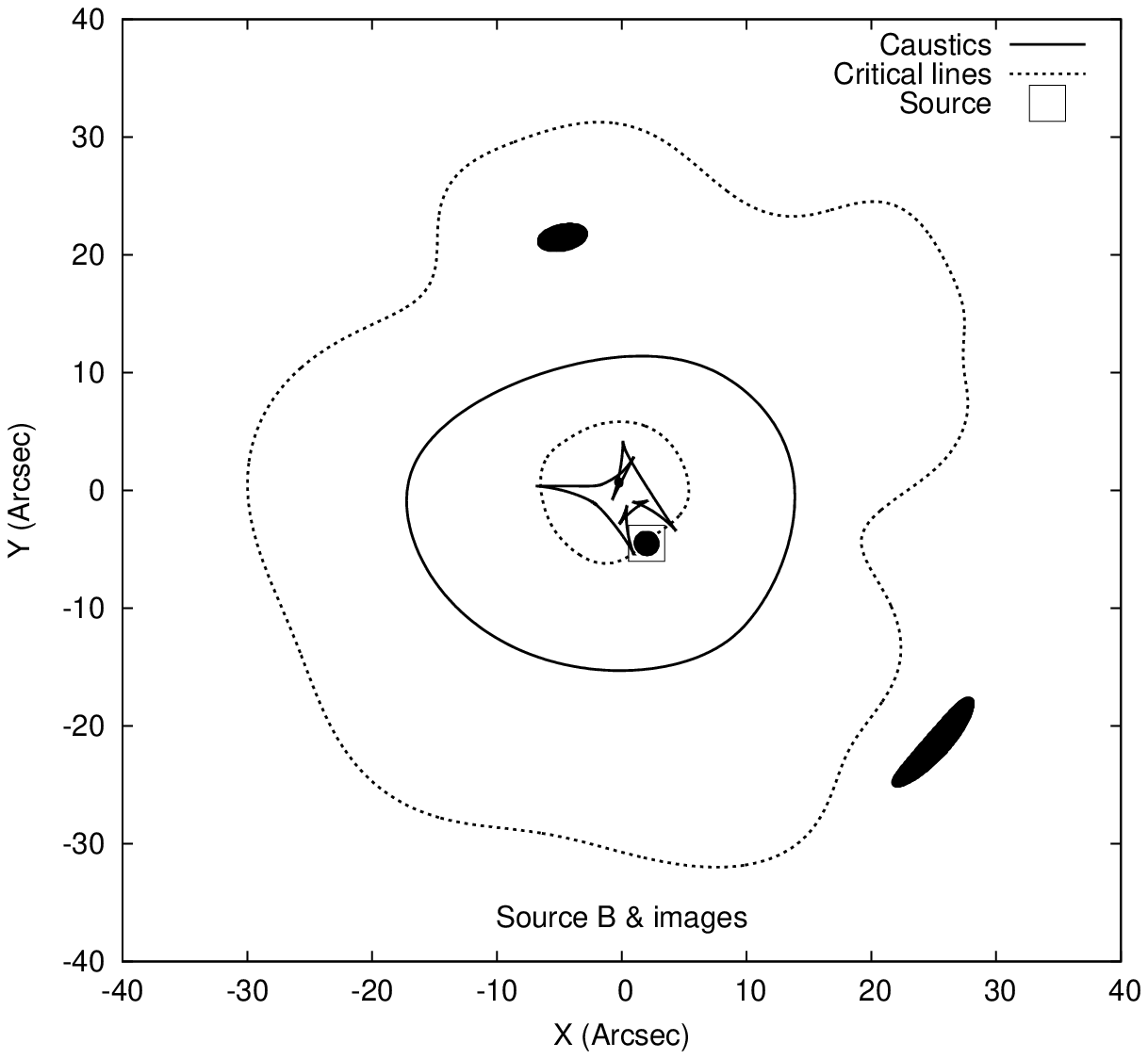}}
			\caption{Left column:~this gravitational lensing situation is used as a starting point to illustrate the
			         general mass sheet degeneracy in which different sources are scaled with different values. The top
					 figure depicts the elliptical mass distribution of the gravitational lens; the two other figures
					 show how this lens creates three images of two different sources. In this example, only the source 
					 in the middle panel (source $A$) will be rescaled, the source in the bottom figure (source $B$) 
					 will not. 
					 Center column:~after modifying the mass distribution in the left column using the method of the
					 monopole basis functions (see text), a rescaled version of source $A$ is needed to create the
					 same images as before. Source $B$ on the other hand does not need to be altered to recreate the
					 same image configuration. To illustrate that the source plane of source $A$ has been rescaled toward
					 the origin can be seen by comparing the new source position to the original one.
					 Right column:~similar to the center column, but in this case a similar interpolation method for the
					 deflection field was used as in the LensPerfect lens inversion procedure \citep{LensPerfect}.}
			\label{fig:sheetmostgeneral}
		\end{figure*}

		The effect is the same as for the original mass sheet degeneracy: the
		images will now correspond to sources which are rescaled by the same factor $\lambda$.
		The magnifications are therefore again modified by $\lambda^{-2}$. The effect
		on the time delay is not as straightforward however. For a circularly symmetric 
		lens, combining equations (\ref{eq:lenspotential}) and (\ref{eq:alphasymm}) shows that
		\begin{equation}
			\psi(\theta) \propto \int_0^\theta \frac{M(\theta')}{\theta'}d\theta' \mcm
		\end{equation}
		i.e. the lens potential, and therefore the time delays, will depend on the precise
		manner in which the interpolation between the two curves in Fig.~\ref{fig:doublesheet}
		is done.

		Using the generalization of the mass sheet degeneracy that was just shown,
		the sources are still rescaled with the same factor. But it is also
		possible to construct a degenerate solution which allows one to scale
		each source with a different factor. 
		To illustrate this, consider the situation shown in the left panel
		Fig.~\ref{fig:sheetmostgeneral}: an elliptical mass
		distribution causes two sources to produce three images each. In what
		follows, only source $A$ will be rescaled with a certain factor. In
		a subsequent step the other source can be rescaled in a similar way,
		using a different factor.
		
		Focusing on only one source and temporarily
		forgetting about the other, the source can be rescaled using the
		classic mass sheet degeneracy from equation~(\ref{eq:sheetdegen}).
		This means that to the initial $\Sigma_0$, a mass distribution $\Sigma_{\rm add}$
		was added to obtain $\Sigma_1$:
		\begin{equation}
			\Sigma_{\rm add}(\Vec{\theta}) 
		                              = (1-\lambda)\left[\Sigma_{\rm cr}(z_{\rm A})-\Sigma_0(\Vec{\theta})\right] \mpt
		\end{equation}
		Taking the second source back into account, it is clear that adding
		$\Sigma_{\rm add}$ to the original mass distribution will cause the images
		of source $A$ to correspond to a scaled version of that source, but
		the effect on the second source is less desirable. 
		
		However, suppose it is possible to modify $\Sigma_{\rm add}$ so that the
		deflection it causes for source $A$ is unaltered, and no deflection is
		caused at the location of the images of source $B$. In that case, after
		adding $\Sigma_{\rm add}$, the images of source $A$ would correspond to a 
		scaled version of that source, and the images of source $B$ would still
		be projected onto the same source area. Calling $I_{\rm A}$ and 
		$I_{\rm B}$ the points of which respectively the images of sources $A$ and
		$B$ consist, $\Vec{\hat{\alpha}}_{\rm add}(I_{\rm B}) = 0$ by equations
		(\ref{eq:lenspotential}) and (\ref{eq:convergence}) 
		implies that $\Sigma_{\rm add}(I_{\rm B}) = 0$.
		If one then constructs $\Sigma_1 = \Sigma_0 + \Sigma_{\rm add}$, in general one 
		finds that 
		\begin{eqnarray}
			\Sigma_1(I_{\rm A}) & = & \lambda\Sigma_0(I_{\rm A}) + (1-\lambda)\Sigma_{\rm cr}(z_{\rm A}) \nonumber \\
			\Sigma_1(I_{\rm B}) & = & \Sigma_0(I_{\rm B}) \label{eq:almostgeneralsheetdegen}
		\end{eqnarray}
		Starting from $\Sigma_1$, in a next step one can then exchange the roles 
		of the two sources and rescale source $B$ with a different factor.

		There still remains the question of how to modify $\Sigma_{\rm add}$ so
		that it does not produce a deflection at the location of the images
		of source $B$. This can be done by using a large number of the monopole
		basis functions we met earlier. If none of these overlap with the
		images of source $A$, the deflection there will be unaltered. To make
		sure that the deflection angle at the location of the images of
		source $B$ vanishes, appropriate weights of these basis functions must
		be sought. An example of this approach can be seen in the center panel
		of Fig.~\ref{fig:sheetmostgeneral}. In this example, the weights were 
		determined by a genetic algorithm, an optimization strategy
		inspired by natural selection. A similar procedure is used as in
		\citet{Liesenborgs4}: a trial solution consists of a set of weights
		for the monopole basis functions. Starting from randomly initialized
		trial solutions, the algorithm then `breeds' increasingly better trial
		solutions by combining existing solutions and introducing mutations. A key
		step in this procedure is the application of selection pressure: better
		solutions should produce more offspring. In this case, a trial solution is 
		deemed better than another simply when its deflection field differs less
		from the desired one at the locations of the images of source $B$. Since, 
		when generating a degenerate solution in this fashion, one still uses a 
		rescaled version of the original mass distribution, this procedure still 
		produces a mass map which appears rescaled, as can be seen in the center 
		panel of Fig.~\ref{fig:sheetmostgeneral}.

		Another approach which does not cause such an apparent rescaling, can be used 
		as well. From the discussion above,
		it is clear that the final version of the mass distribution $\Sigma_{\rm add}$
		should produce a specific value of the deflection angle $\Vec{\hat{\alpha}}_{\rm add}$
		at the location of the images of source $A$. On the other hand, at the location
		of the images of source $B$, no deflection should be produced. We therefore
		know the deflection angle at specific locations, and at other locations
		the deflection angle can take a large number of values, the only real constraint
		being that no additional images should be produced. Interpolating a deflection
		field in such a way that $\Vec{\nabla}\times\Vec{\hat{\alpha}}_{\rm add} = 0$, or equivalently
		that the deflection originates from a scalar lensing potential, is
		precisely what is done in the LensPerfect lens inversion procedure \citep{LensPerfect}.
		Using a similar interpolation procedure to determine $\Vec{\hat{\alpha}}_{\rm add}$ and
		therefore $\Sigma_{\rm add}$, yields the result shown in the right panel
		of Fig.~\ref{fig:sheetmostgeneral}. While more difficult to ensure that the 
		overall mass distribution is positive, in this example it produces an interesting
		result. Comparing the resulting mass distribution with the original one in the
		left panel, one sees that the differences are rather subtle. And since no rescaling
		of the mass density itself seems to be involved, one would not suspect a mass sheet-like
		degeneracy to be at work here.

		Note that in principle, a similar construction is possible when only using
		point sources. When the deflection angles are only conserved at very specific
		point locations and not over a more extended area, the gradients of the deflection
		field will change and therefore also the surface density and the magnification
		factors. In particular, the relation in equation (\ref{eq:almostgeneralsheetdegen})
		will no longer hold. However, since this no longer corresponds to a rescaling of the
		sources, it can hardly be considered as caused by the mass sheet degeneracy.

			\begin{figure*}
				\centering
				\subfigure{\includegraphics[width=0.32\textwidth]{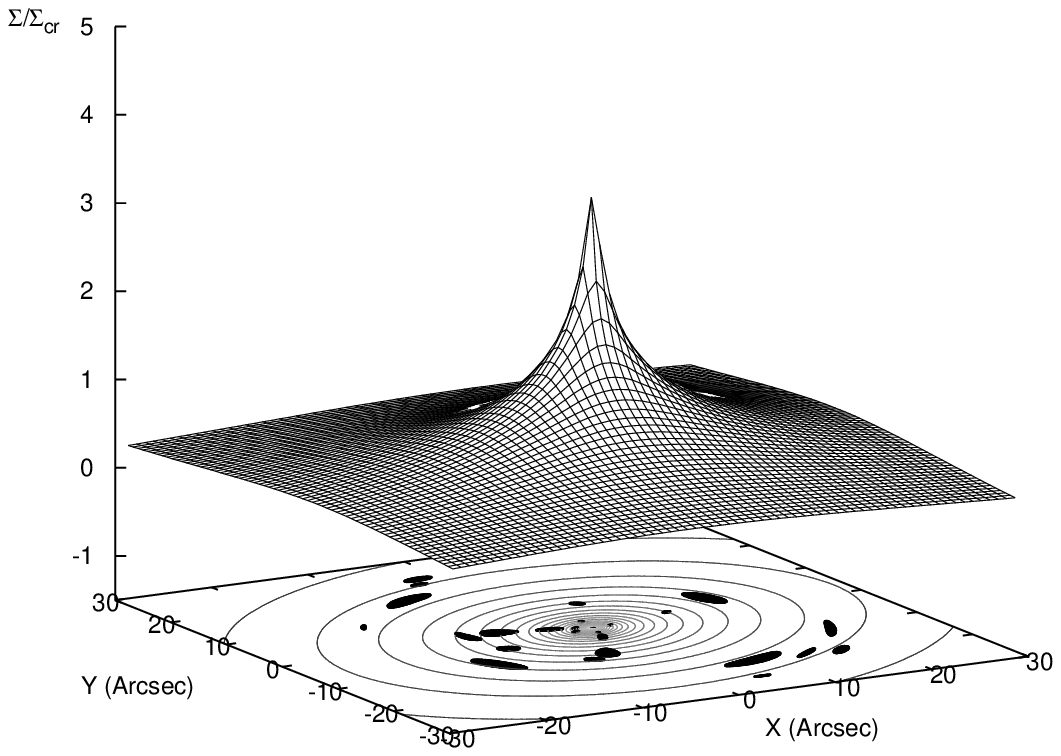}}
				\subfigure{\includegraphics[width=0.32\textwidth]{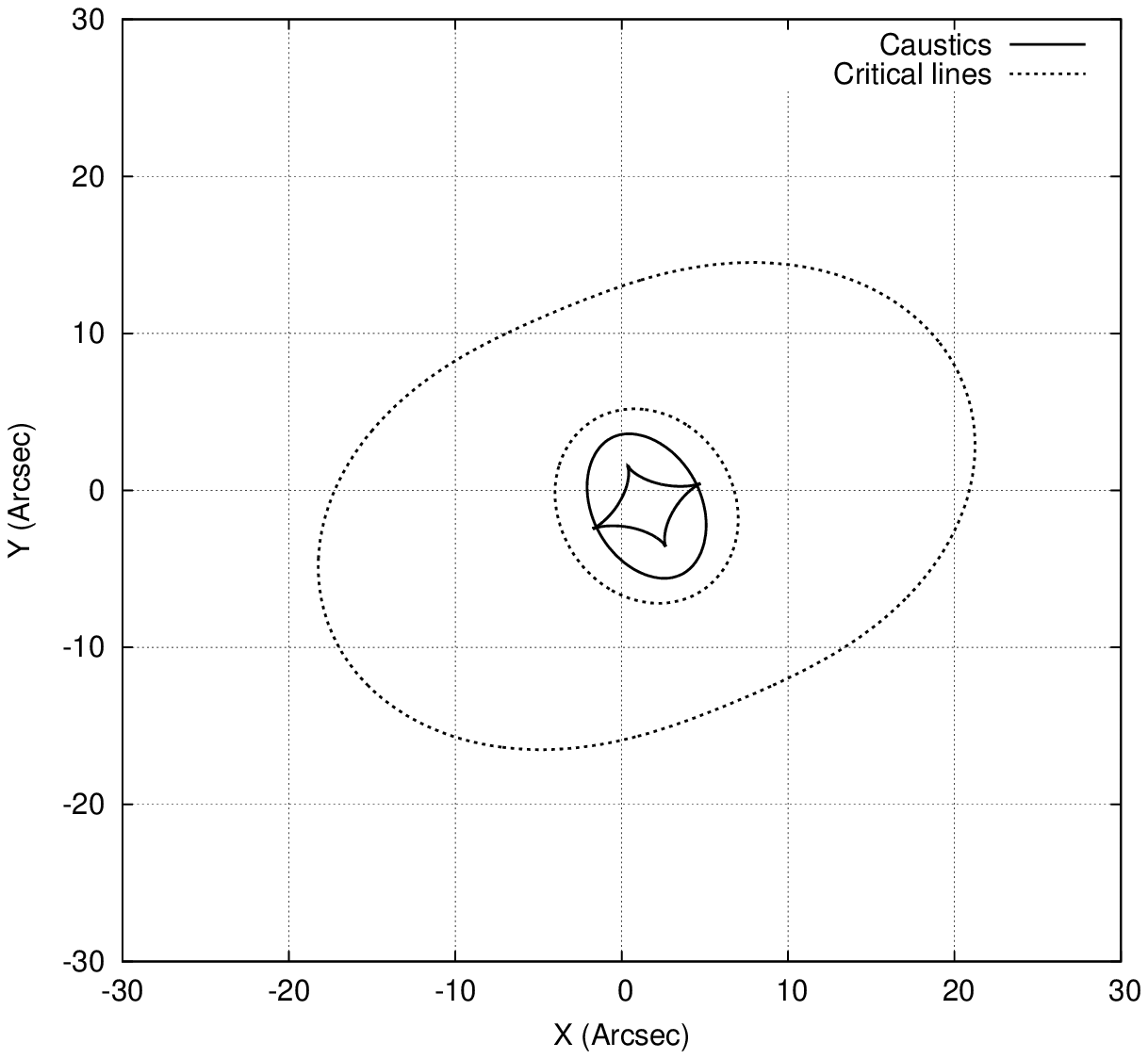}}
				\subfigure{\includegraphics[width=0.32\textwidth]{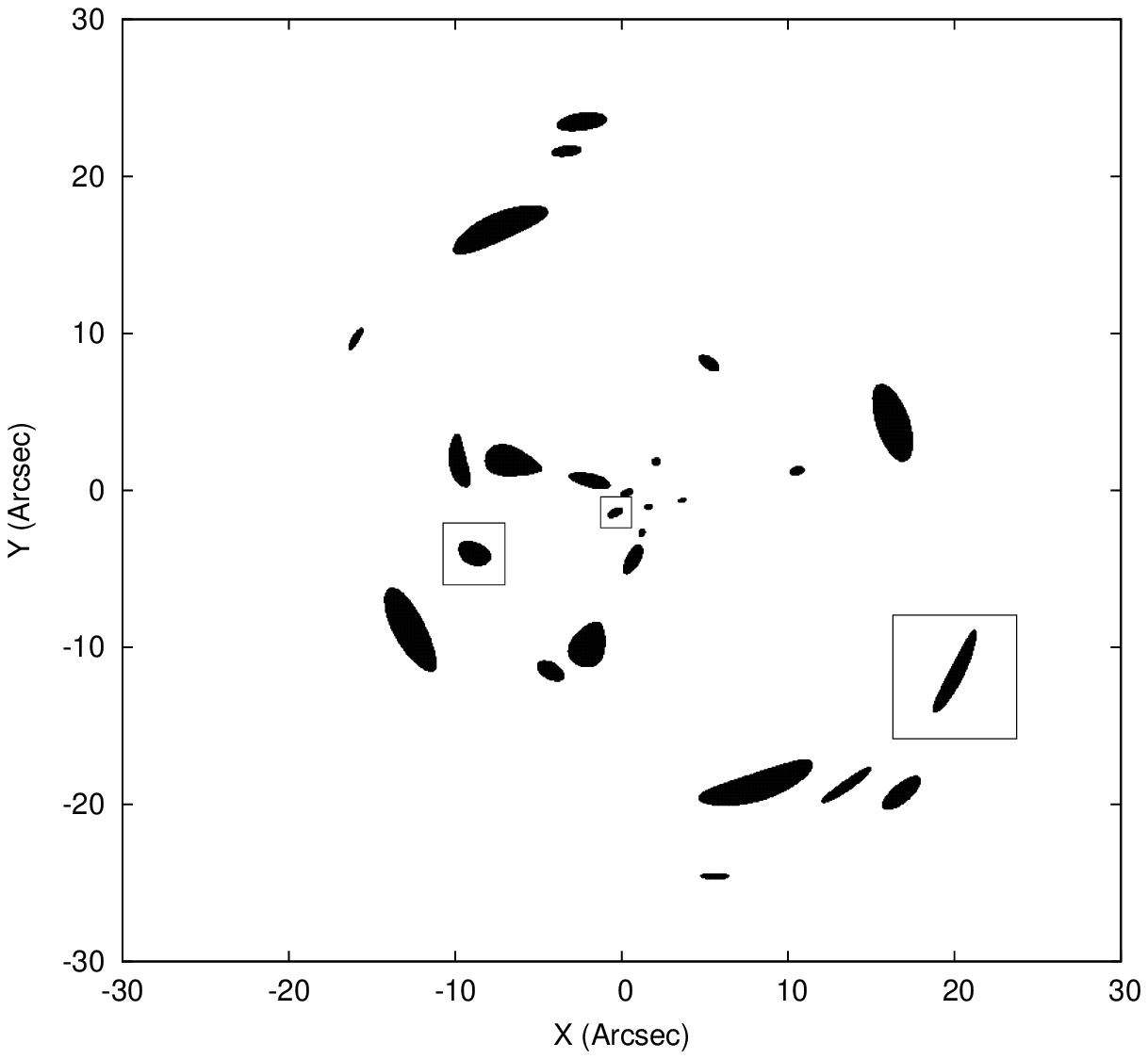}}
				\caption{Left panel:~the input mass distribution for a simulated lensing
					scenario consists of an elliptical generalization of a NFW profile. Eight
					distinct sources generate 26 images which are shown on top of the contour
					map.
					Center panel:~for a source at $z = 3$, the mass distribution from the left
					panel generates these critical lines and caustics.
					Right panel:~the same 26 input images as in the left panel are shown here.
					The three images of a single source which will be used for additional
					time delay constraints later, are surrounded by squares.}
				\label{fig:experimentsetup}
			\end{figure*}

			\begin{figure*}
				\centering
				\subfigure{\includegraphics[width=0.32\textwidth]{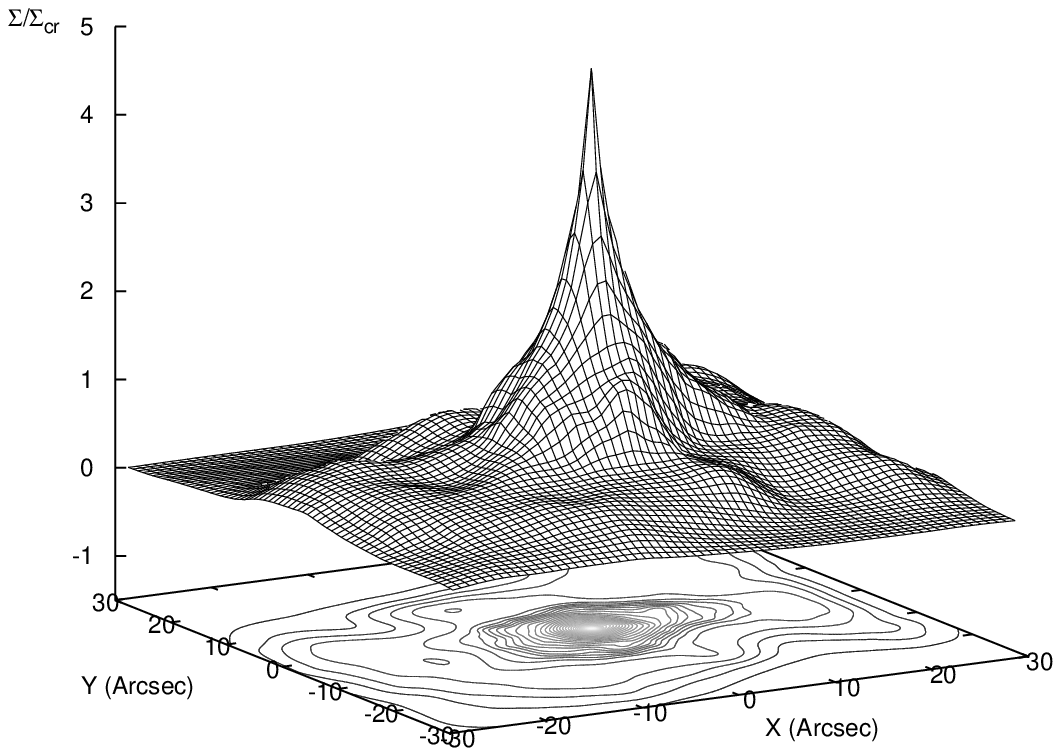}}
				\subfigure{\includegraphics[width=0.32\textwidth]{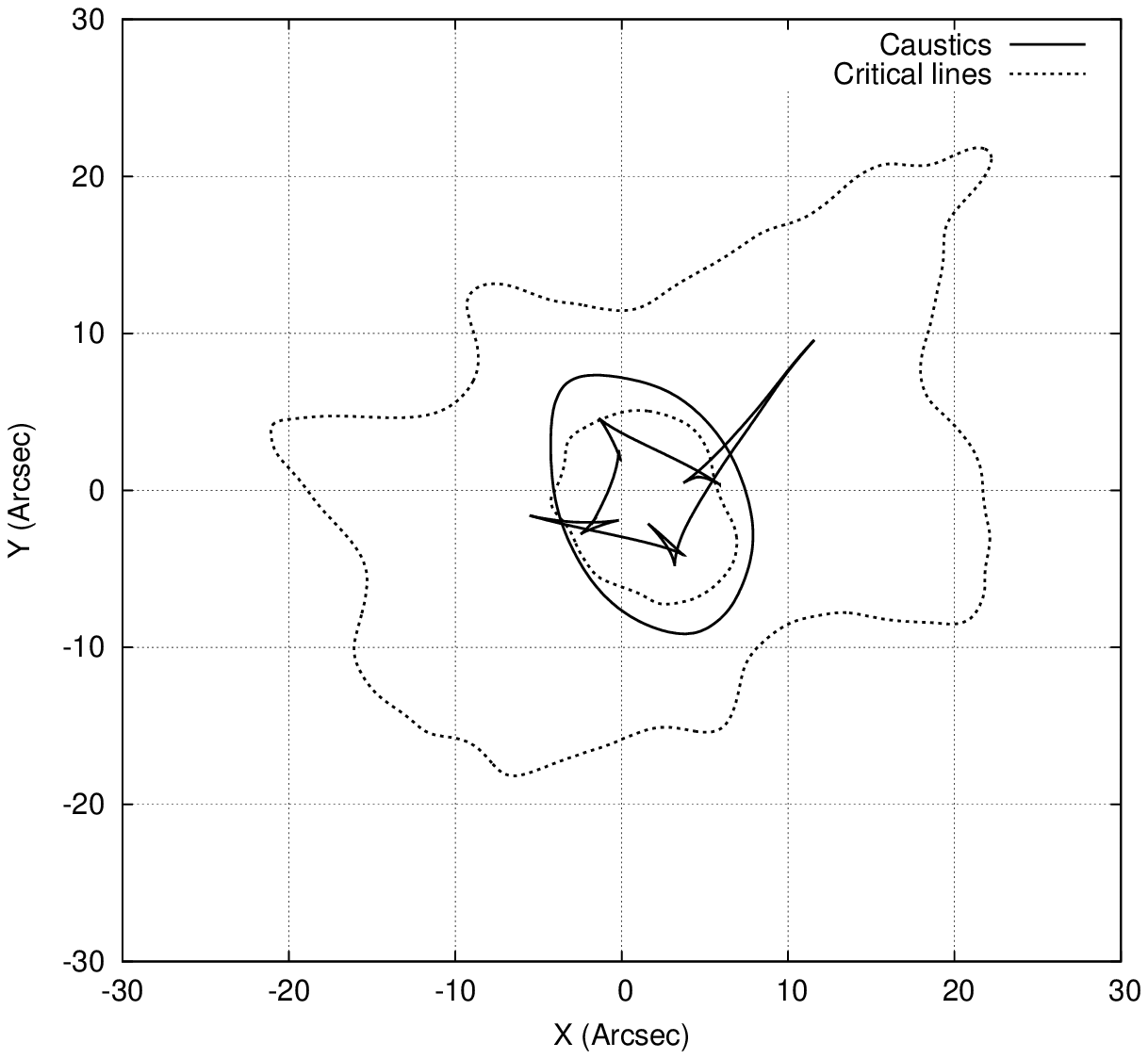}}
				\subfigure{\includegraphics[width=0.32\textwidth]{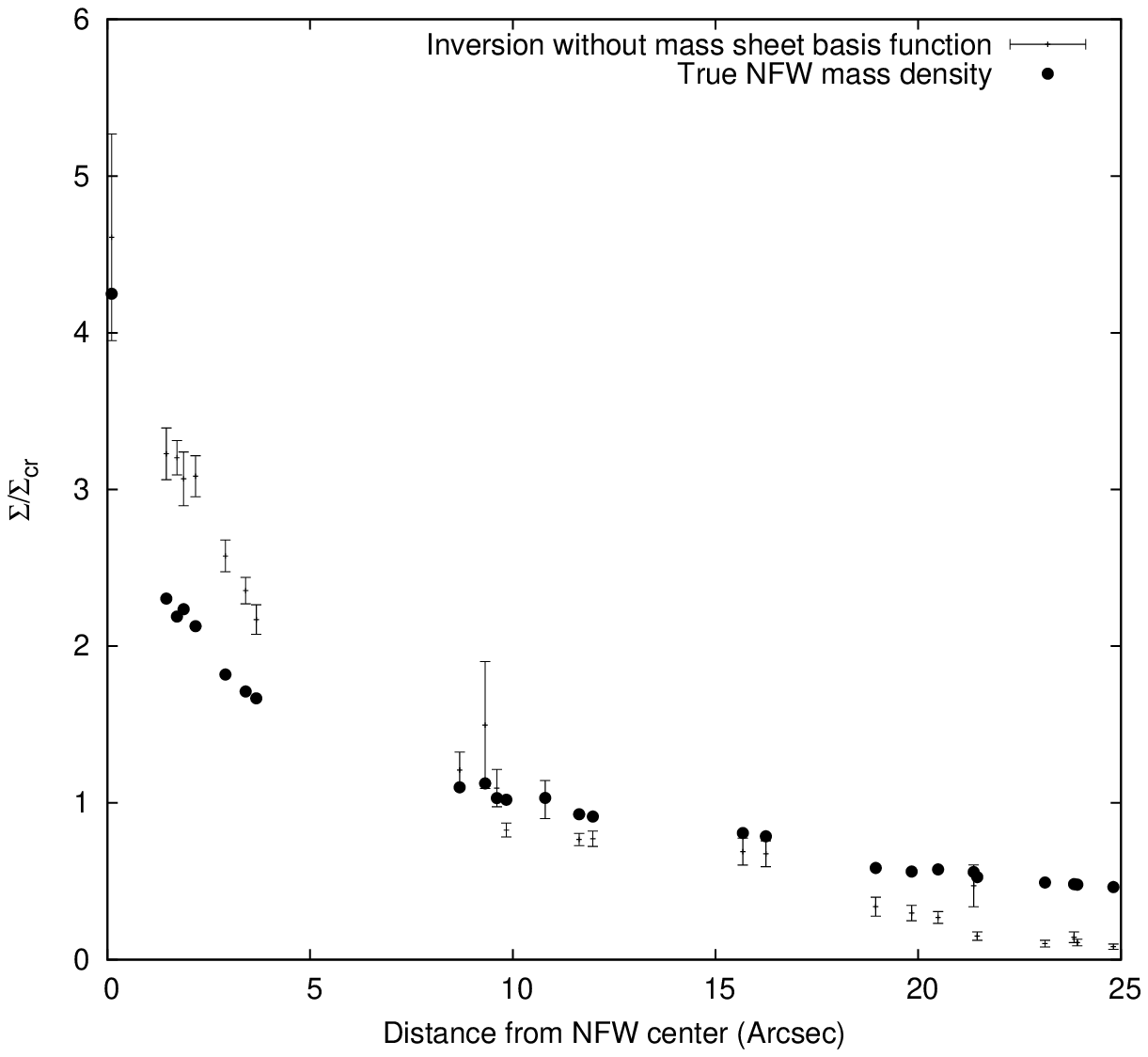}}
				\caption{Left panel:~when only positional information about the image systems is
					used and only Plummer basis functions are used, corresponding to the
					algorithm as described in \citet{Liesenborgs}, the average of 20 individual
					solutions yields this mass distribution. Although the overall shape is
					similar to the mass distribution in Fig.~\ref{fig:experimentsetup}, this
					mass map is clearly much more steep.
					Center panel:~the critical lines and caustics for this average solution.
					Comparing this to the center panel of Fig.~\ref{fig:experimentsetup}, one
					immediately notices that the scale of the caustics is much larger, indicating
					the presence of a mass sheet-like degeneracy.
					Right panel:~the densities of the individual solutions are compared to the
					true input mass densities, precisely at the locations of the image centers.
					Being much more steep and apparently lacking a constant offset again reveals
					the presence of a mass sheet-like degeneracy.
					}
				\label{fig:invnosheet}
			\end{figure*}

	\section{Experiment}\label{sec:experiment}

		\subsection{Positional information}

			To further illustrate the significant practical problems posed
			by these degeneracies, we shall use a simulated lensing scenario. 
			The left panel of Fig.~\ref{fig:experimentsetup} shows the true mass map
			used in the simulation, an elliptical generalization of a NFW mass 
			distribution \citep{1996ApJ...462..563N,2000ApJ...534...34W}, with concentration parameter
			$c_{\rm vir} = 5$ and scale parameter $r_{\rm s} = 50$ arcseconds. The lens
			was placed at a redshift of $0.7$ in a cosmology with $H_0 = 70$ km s$^{-1}$ Mpc$^{-1}$,
			$\Omega_{\rm m} = 0.3$ and $\Omega_\Lambda = 0.7$. The surface density is expressed
			in terms of the critical density for a source at $z = 3$. Eight sources were placed
			at different redshifts. These generate 26 images which can be seen in the contour map,
			and also in the right panel of the same
			figure. The redshifts of the sources lie between $z = 2.7$ and $z = 3.4$.
			The center panel of Fig.~\ref{fig:experimentsetup} shows the critical lines and
			caustics for a source at $z = 3$. We shall be interested in comparing the scale
			of the caustics here to that of reconstructions, as rescaling the source plane
			is the hallmark of the mass sheet degeneracy.

			First, a reconstruction is done using the algorithm presented in \cite{Liesenborgs}.
			In this method, Plummer \citep{1911MNRAS..71..460P} basis functions are arranged
			according to an adaptive grid, and the weights of these basis functions are determined
			using a genetic algorithm. The left panel of Fig.~\ref{fig:invnosheet} shows the
			resulting mass distribution, an average of 20 individual reconstructions. Overall, 
			it has a similar appearance as the input mass
			map shown in Fig.~\ref{fig:experimentsetup}, and its critical lines (center panel) do
			bear resemblance to those of the true lens. The resulting mass distribution, however,
			is considerably steeper than the one used in the simulation, which can also clearly
			be seen in the right panel of Fig.~\ref{fig:invnosheet}. In this image we show the densities
			measured at the center of each image. For the sake of visualization, instead of using
			the two dimensional coordinate of each image center point, the distance to the NFW
			center is used on the $x$-axis. Comparing the two profiles on that figure, one is 
			immediately reminded of the mass sheet degeneracy, i.e. the profiles seem to differ
			mostly by a rescaling and the addition of a specific offset. Similarly, when looking
			at the caustics in the center panel of Fig.~\ref{fig:invnosheet}, they seem to be
			an expanded version of the ones of the input lens, again a feature that hints at the
			presence of the mass sheet degeneracy.

			Due to the inherent uncertainty in between the images caused by the monopole degeneracy,
			the most sensible thing to do in trying to retrieve the NFW parameters, is to perform a fit
			based on the densities precisely at the locations of the images. When looking at the 
			steepness of the reconstructed mass map, it will come as no surprise that extracting
			the concentration and scale parameters from the reconstructed densities will not yield
			accurate results. In this particular case, the estimates are $c_{\rm vir} = 14.1^{+0.6}_{-0.5}$
			and $r_{\rm s} = 11.7^{+0.6}_{-0.7}$ arcseconds respectively (68\% confidence level).

			\begin{figure*}
				\centering
				\subfigure{\includegraphics[width=0.32\textwidth]{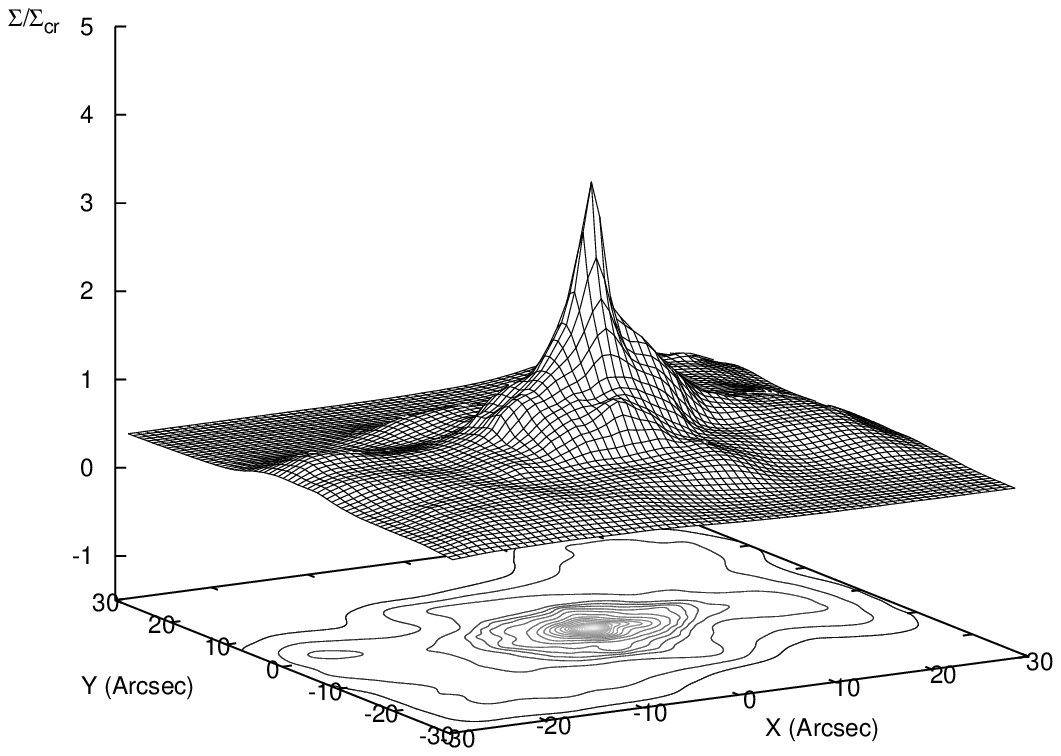}}
				\subfigure{\includegraphics[width=0.32\textwidth]{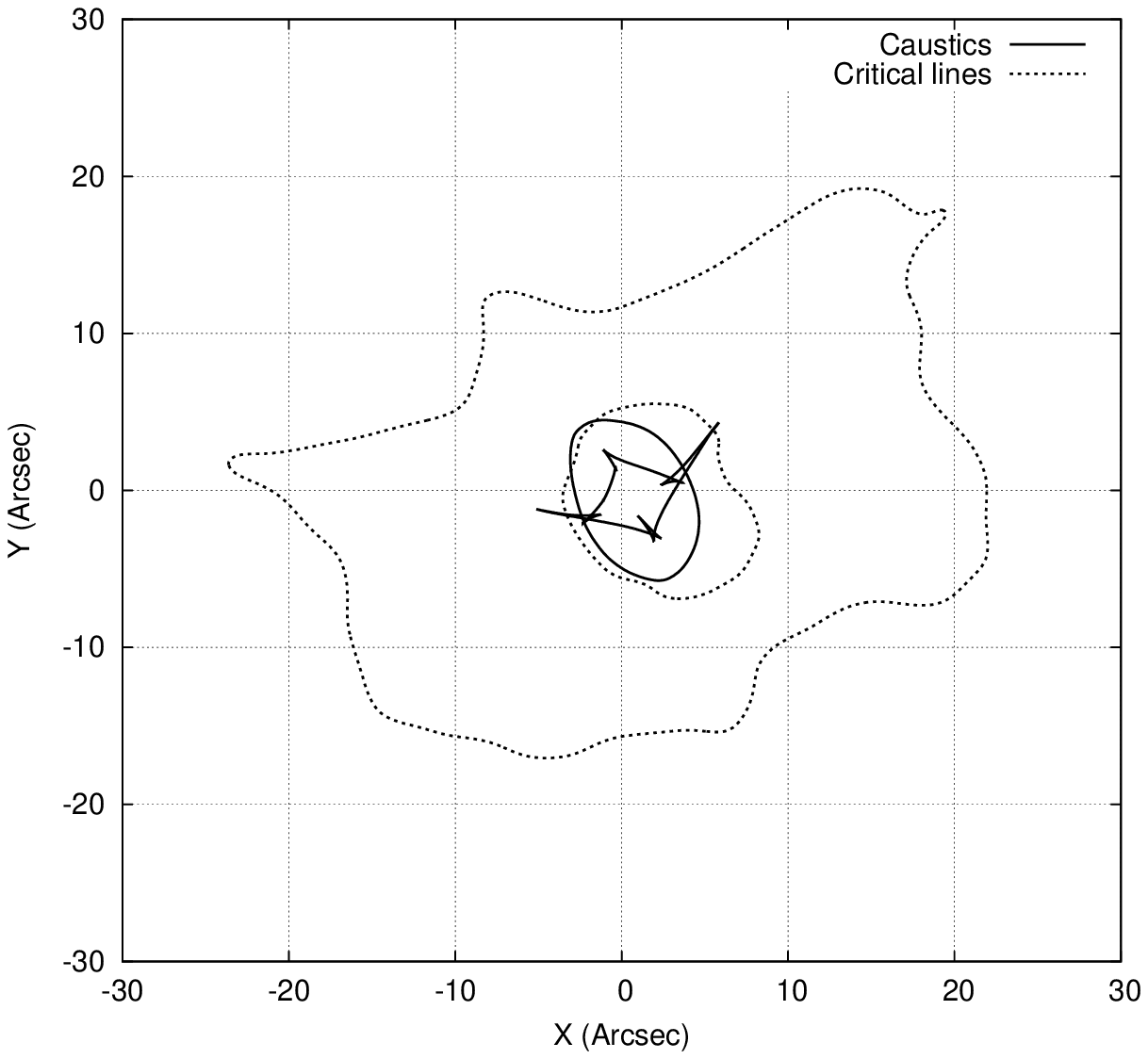}}
				\subfigure{\includegraphics[width=0.32\textwidth]{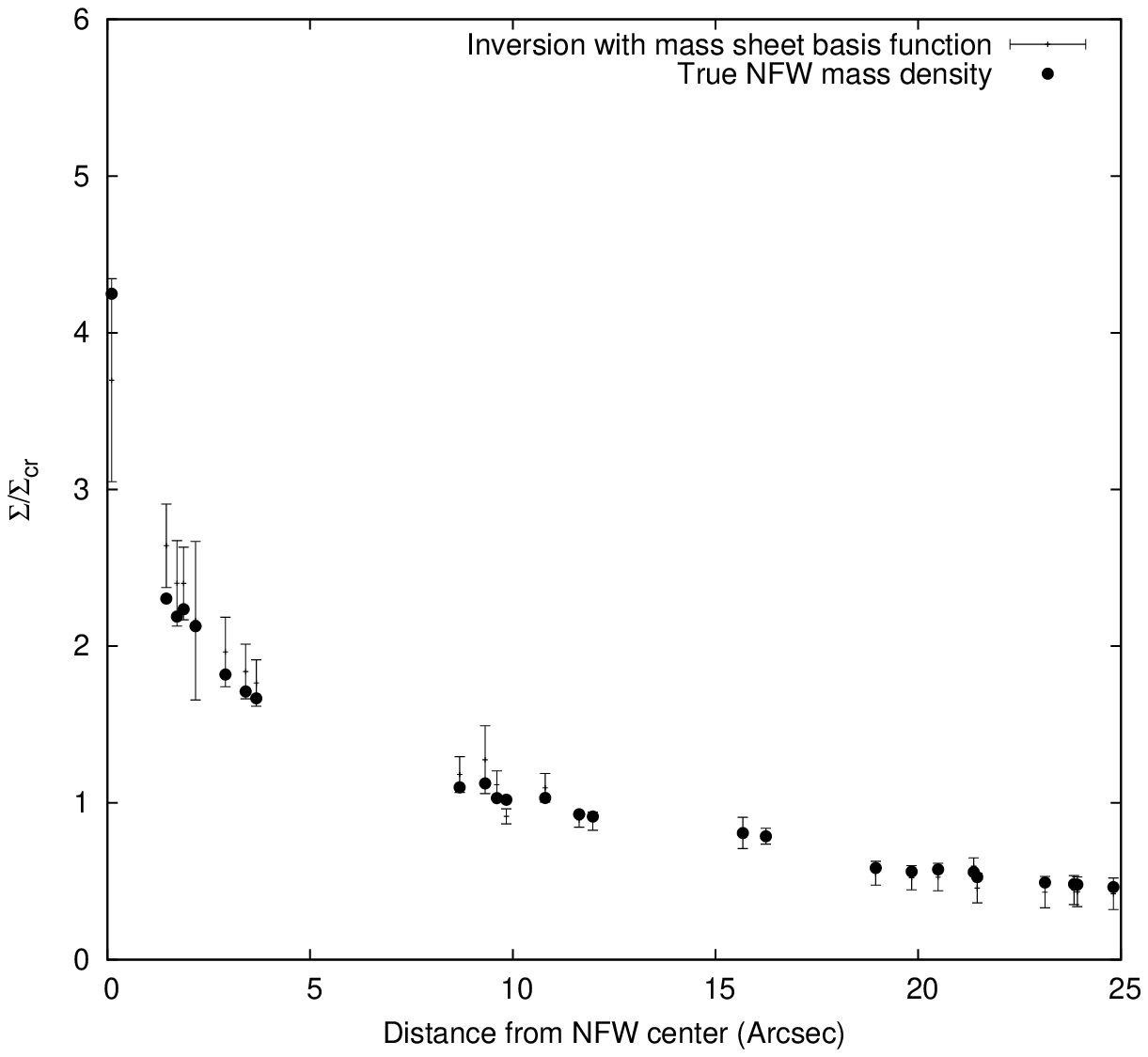}}
				\caption{Left panel:~to generate this average solution, the same algorithm was
					used as in Fig.~\ref{fig:invnosheet}, but this time a constant density sheet
					was allowed as a basis function as well (see \citet{Liesenborgs5}). This
					solution resembles the input mass distribution more closely.
					Center panel:~when the critical lines and caustics for this solution are
					calculated, the source plane no longer exhibits an obvious rescaling
					compared to the input situation.
					Right panel:~comparing the densities at the image centers to those of the
					input mass map confirms that this reconstruction is much more similar to
					the true situation than the mass distribution from Fig.~\ref{fig:invnosheet}.}
				\label{fig:invsheet}
			\end{figure*}

			Using Plummer basis functions it is of course rather difficult to mimic the effect
			of a sheet of mass. Given the fact that within the images a considerable mass sheet-like
			component seems to be present in the input lens, the fact that the reconstruction is rather
			different is not that surprising. In a second reconstruction not only Plummer basis functions
			were used, but a single mass sheet basis function was present as well, as is also explained
			in \citet{Liesenborgs5}. The resulting mass distribution, again an average of 20 individual
			solutions, can be seen in the left panel of Fig.~\ref{fig:invsheet}. This mass distribution
			is clearly a lot more similar to the input mass map, also obvious from the right
			panel of the same figure. As can be expected from the similarity between the true and reconstructed
			lenses, the caustics shown in the center panel no longer show a clear difference in scale.
			When performing a fit in a similar fashion as the previous reconstruction, one finds NFW
			parameters of $c_{\rm vir} = 5.9^{+0.6}_{-0.5}$ and $r_{\rm s} = 38.3^{+4.9}_{-4.4}$ arcseconds,
			corresponding better to the true values.

			Unfortunately, in this example there is no real reason to prefer one reconstruction over the
			other, meaning that apart from the overall shape of the true mass distribution, little
			would be revealed by performing these inversions. Furthermore, in both cases fitting the
			NFW shape yields an overestimate of the concentration parameter and an underestimate of
			the scale radius, although the effect is far more serious in the first inversion.
	
			\begin{figure*}
				\centering
				\subfigure{\includegraphics[width=0.32\textwidth]{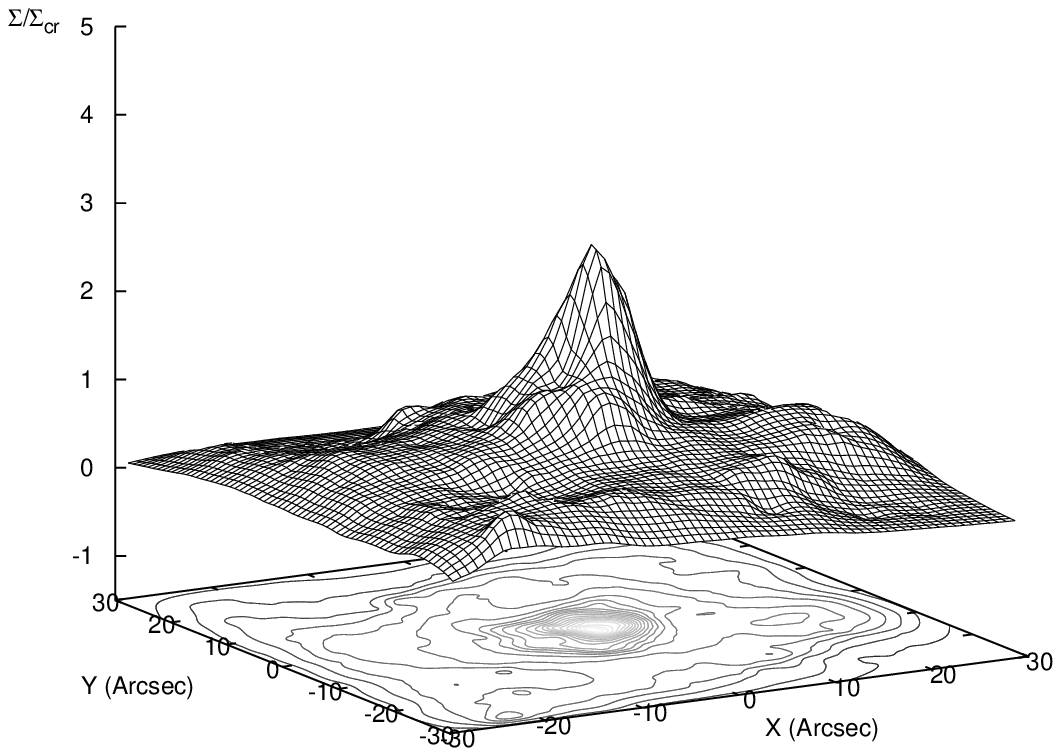}}
				\subfigure{\includegraphics[width=0.32\textwidth]{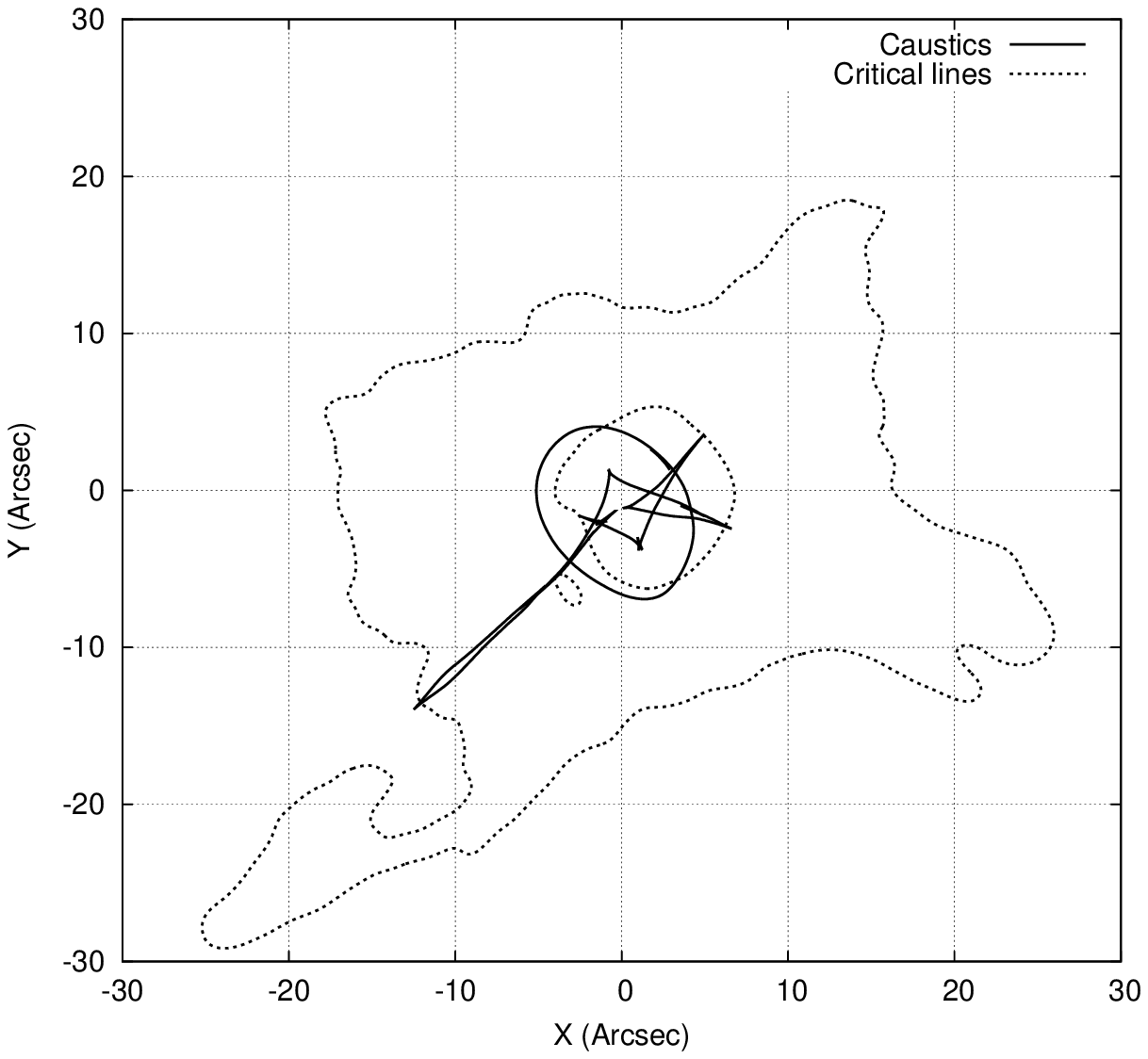}}
				\subfigure{\includegraphics[width=0.32\textwidth]{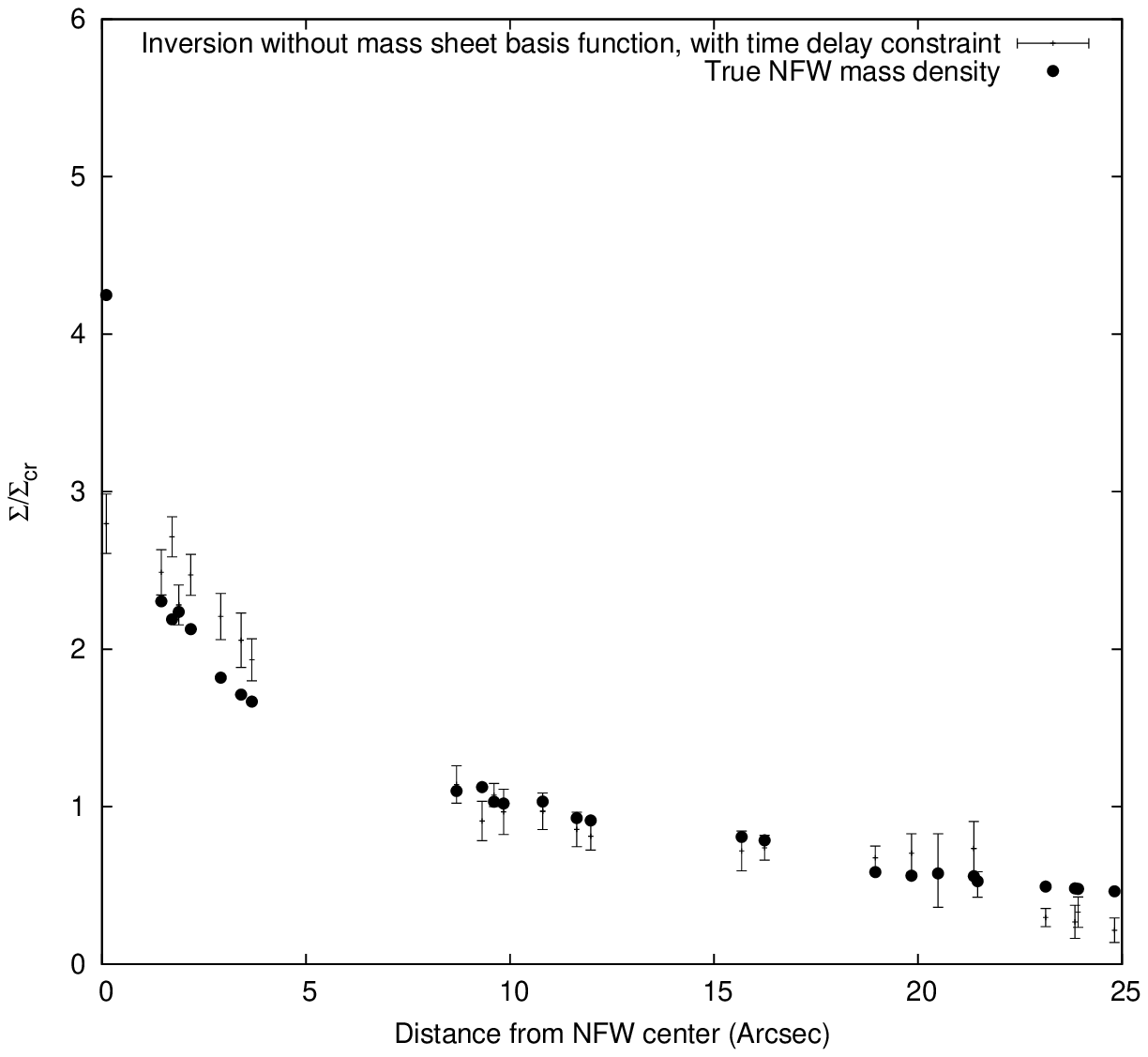}}
				\caption{Left panel:~this is the average solution when only Plummer basis functions
					are used, similar to Fig.~\ref{fig:invnosheet}, but as an additional constraint
					the time delays between the images indicated in the right panel of Fig.~\ref{fig:experimentsetup}
					were included. Comparing this result to the one in Fig.~\ref{fig:invnosheet}
					reveals that the addition of this information has a profound effect on the
					steepness of the solution.
					Center panel:~thanks to the presence of the time delay information the caustics 
					of this average solution no longer display the obvious rescaling that could be
					observed earlier.
					Right panel:~as was already suggested by the left panel, comparing the densities
					at the image locations shows that the time delay information for even a single
					image system seems to pull all reconstructed densities towards the true densities.}
				\label{fig:invnosheettd}
			\end{figure*}
	
			\begin{figure*}
				\centering
				\subfigure{\includegraphics[width=0.32\textwidth]{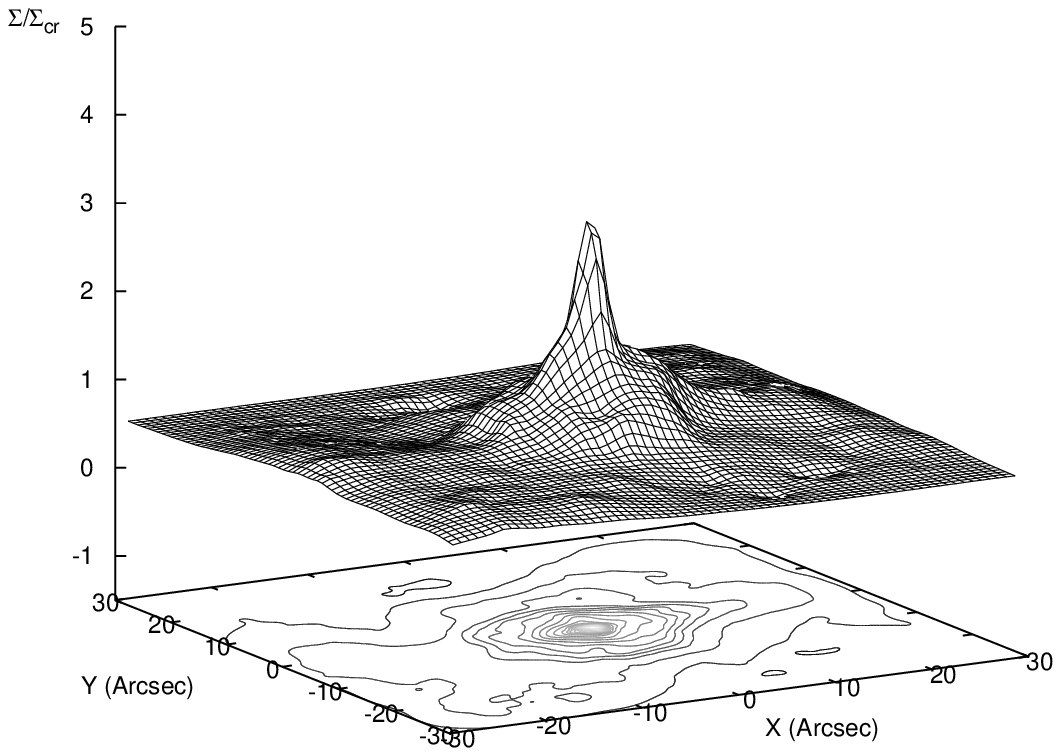}}
				\subfigure{\includegraphics[width=0.32\textwidth]{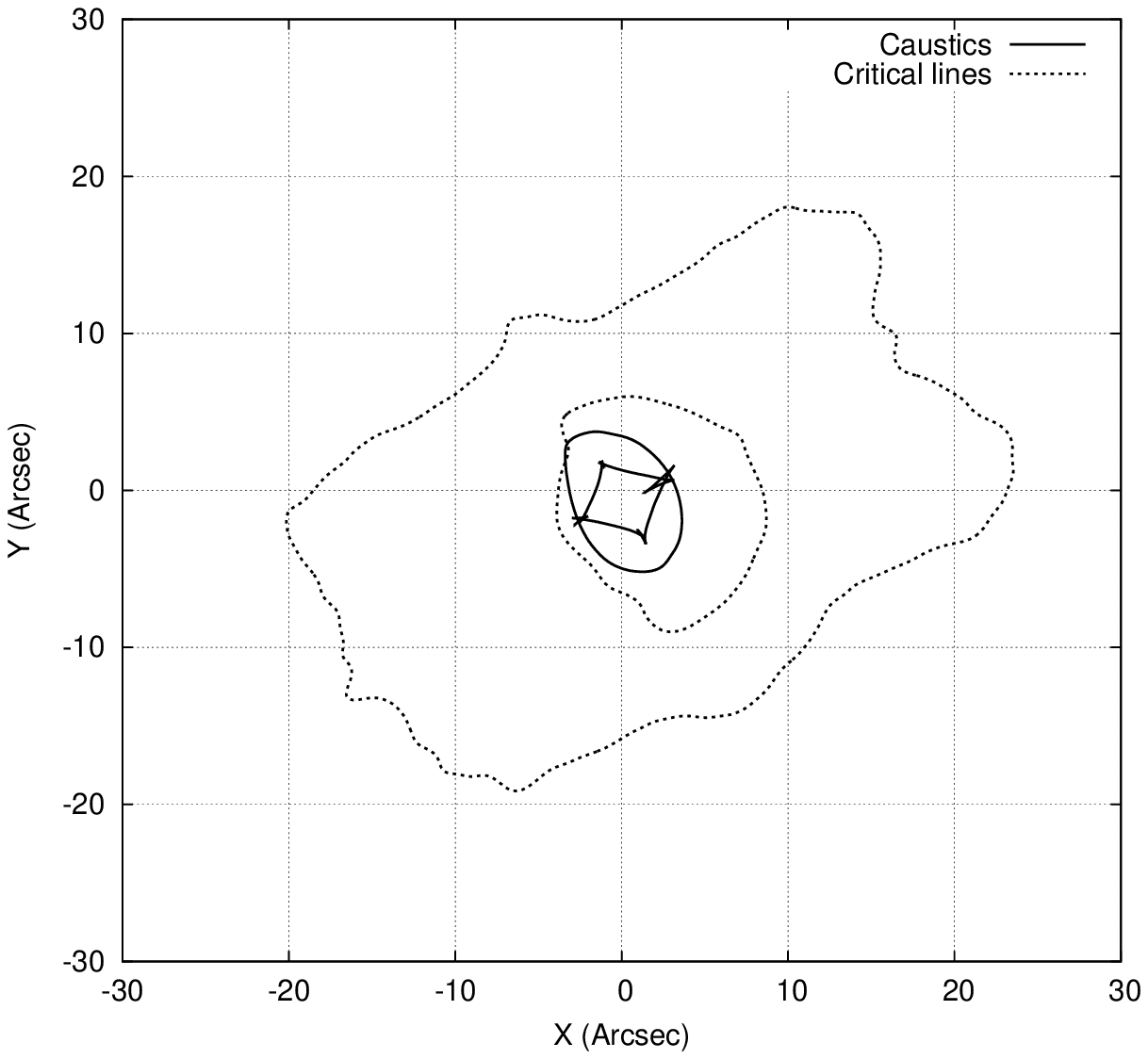}}
				\subfigure{\includegraphics[width=0.32\textwidth]{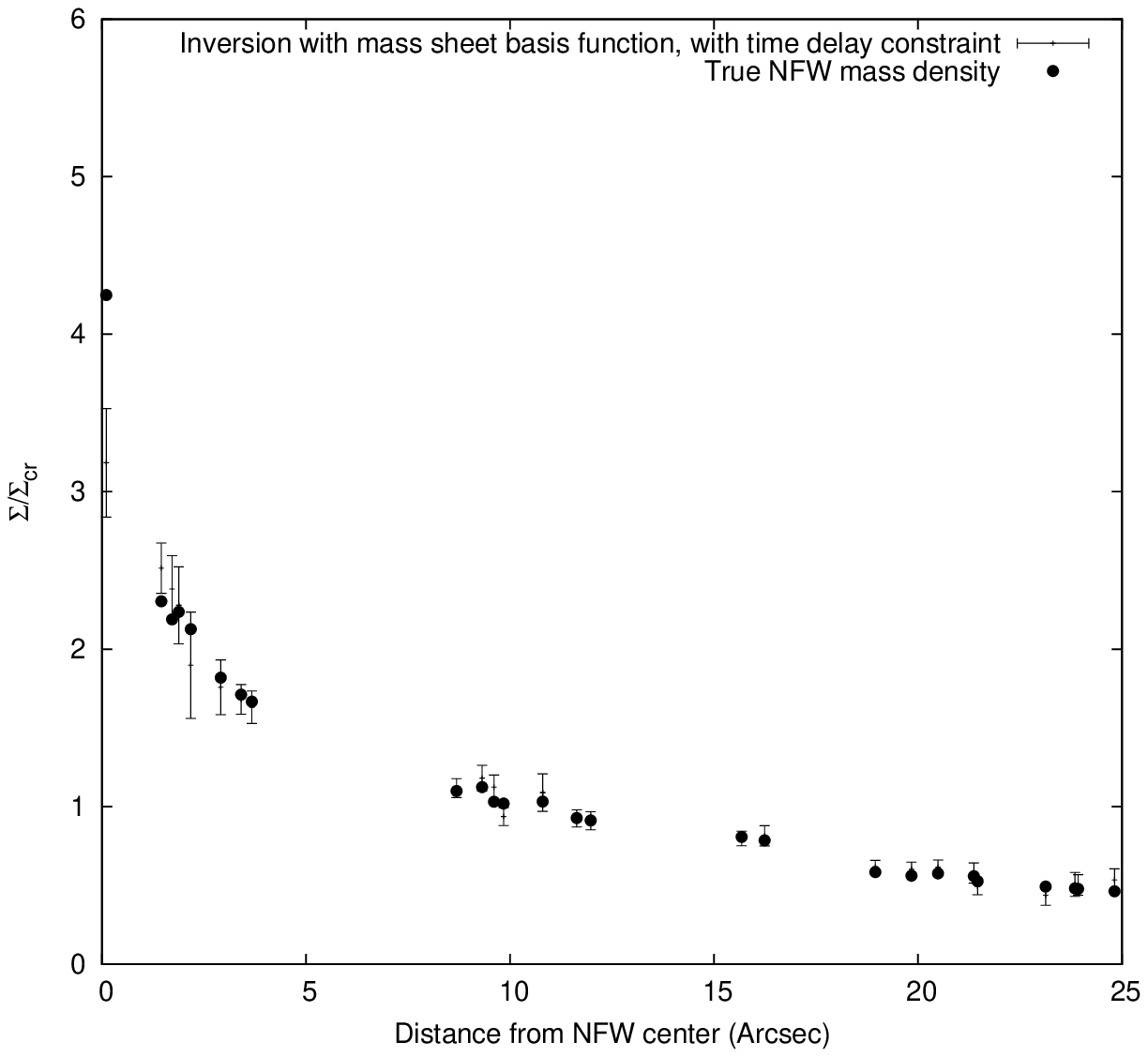}}
				\caption{Left panel:~when both time delay information about the system marked in the
					right panel of Fig.~\ref{fig:experimentsetup} and a mass sheet basis function
					are allowed, this average result is obtained. Apart from the sharp central peak,
					the mass distribution is very similar to the input mass map.
					Center panel:~calculating the critical lines and caustics of this solution reveals
					that the source plane is very similar in scale to the one of the input lens.
					Right panel:~in this case the densities at the image locations are extremely similar
					to the input densities. Only the density at the point closest to the center of the
					input NFW distribution appears to be underestimated systematically.}
				\label{fig:invsheettd}
			\end{figure*}

		\subsection{Including time delay information}

			Earlier in this article it was mentioned that time delay information can help
			not only in breaking the most basic mass sheet degeneracy; also for the
			generalized versions time delays depend on the precise construction method of
			the degenerate mass map. In \citet{Liesenborgs5} a method was described with
			which time delay information can be added as a constraint, now using a so-called
			multi-objective genetic algorithm. 
			To explore the value of such time delays, time delay information of a single system of images
			was added to the inversion procedure. The three images used for this added constraint 
			are indicated in the right panel of Fig.~\ref{fig:experimentsetup} and correspond 
			to a lensed source at redshift 2.75.

			First, the inversion was run using Plummer basis functions alone, i.e. no
			mass sheet basis function was used. The resulting mass distribution can
			be seen in the left panel of Fig.~\ref{fig:invnosheettd}. Concerning the
			input to the inversion procedure, the only difference with the results
			obtained in Fig.~\ref{fig:invnosheet} is the addition of time delay information.
			The effect of this information is impressive however: especially when
			comparing the right panels of these two figures it is clear that the
			time delay constraint seems to pull the densities at all image locations
			towards the true densities. The caustics in the center panel confirm the
			mass sheet-breaking effect: the overall scale of the source plane seems
			to match the true one more closely. Note that this is solely the effect
			of time delays for a single image system. When determining the NFW parameters
			as before, one now finds $c_{\rm vir} = 7.4^{+0.6}_{-0.5}$ and 
			$r_{\rm s} = 28.8^{+2.8}_{-2.9}$ arcseconds. This result still does not
			reflect the true parameters, but is far less deviant than the estimates
			based on Fig.~\ref{fig:invnosheet}.
			
			As a final inversion, both the time delay information and the mass sheet
			basis function was used, i.e. the inversion is similar to the one reflected
			by Fig.~\ref{fig:invsheet}, but now including time delay information about
			one set of images. The resulting mass map is depicted in the left panel
			of Fig.~\ref{fig:invsheettd}; the densities at the image locations can be
			seen in the right panel of the same figure. The effect of the added time
			delay information is not as impressive as in the previous example, but still
			it leads to an improvement of the estimated densities at most image points.
			Only the density at the point closest to the NFW center seems to be underestimated
			systematically. This single outlier does not seem to have a profound adverse
			effect on the estimation of NFW parameters however, which now are 
			$c_{\rm vir} = 4.8^{+0.4}_{-0.3}$ and $r_{\rm s} = 52.1^{+5.2}_{-4.1}$ arcseconds,
			very close to the true values.

	\section{Discussion and conclusion}\label{sec:conclusions}

		The examples shown in this article make it clear that these lensing
		degeneracies are closely related to substructure in the mass density. The
		monopole degeneracy is the worst, as it has no effect on any property of the
		images: the same sources predict the same images, with identical magnifications
		and identical time delays. In essence it means that we do not have a firm
		handle on the mass distribution in between the images, the only real constraint
		being the absence of unobserved images. This illustrates both the importance of
		gravitational lens systems with a multitude of images and when necessary, of 
		the use of the null space, i.e. the locations where no images are observed.

		In its most general version, the mass sheet degeneracy does not require an
		actual sheet of mass, nor does it necessarily modify the slope of the
		mass distribution, as the alternative name of steepness degeneracy might
		suggest. Comparing equations (\ref{eq:sheetdegen}), (\ref{eq:doublesheetdegen})
		and (\ref{eq:almostgeneralsheetdegen}), it becomes clear that when scaling
		source $A$ with factor $\lambda_{\rm A}$ and source $B$ with factor $\lambda_{\rm B}$,
		the effect on the mass density is the following:
		\begin{eqnarray}
			\Sigma_1\left(I_{\rm A}\right) & = & \lambda_{\rm A} \Sigma_0\left(I_{\rm A}\right) + (1-\lambda_{\rm A}) \Sigma_{\rm cr}\left(z_{\rm A}\right) \nonumber \\
			\Sigma_1\left(I_{\rm B}\right) & = & \lambda_{\rm B} \Sigma_0\left(I_{\rm B}\right) + (1-\lambda_{\rm B}) \Sigma_{\rm cr}\left(z_{\rm B}\right) \label{eq:generalsheetdegen} \mcm
		\end{eqnarray}
		where $I_{\rm A}$ and $I_{\rm B}$ are again the locations of the images of
		source $A$ and $B$ respectively. Of course, this can be generalized to any
		number of sources and images. Apart from rescaling each source in the system,
		equation (\ref{eq:generalsheetdegen}) illustrates that this degeneracy changes
		the density precisely at the locations of the images, i.e. it necessarily
		introduces substructure. The effect on the time
		delays is in general not straightforward implying that time delay information
		may be very valuable in helping to break this degeneracy. The simple effect on the magnification
		is not easy to use since one most often does not have information about the
		size of the source. Efforts to use the magnification information to avoid the
		mass sheet degeneracy have been undertaken however, e.g. in \citet{1998ApJ...501..539T}.

		The fact that both monopole and mass sheet degeneracies are -- at least in
		principle -- possible with any amount of sources and images, combined with
		the fact that they are both linked to substructure in the mass distribution,
		means that in reality it is the prior information on the degree of smoothness 
		of the mass distribution that effectively breaks these degeneracies. The
		degree of smoothness can depend implicitly on the method used, e.g. in
		parametric methods like the LensTool method \citep{2007NJPh....9..447J},
		the method of \citet{2010MNRAS.408.1916Z} which is largely based on the
		distribution of the light, or the method of the authors in which overlapping
		smooth basis functions are used (e.g. \citet{Liesenborgs2}). Other inversion
		procedures use an explicit regularization scheme or an explicit Bayesian 
		prior, e.g. in the PixeLens method (\citet{2004AJ....127.2604S}, \citet{2008ApJ...679...17C}).
		The dependence on explicit or implicit assumptions about the smoothness of
		the solution of course means that one must be careful when
		interpreting obtained inversion results, not only in lensing systems with
		few sources, but also when a larger amount of sources are available, as
		indicated by the experiment. This same experiment also highlights the
		importance of time delay information, which easily asserts a global effect
		on the reconstruction. It is unfortunate that in many cases of interest these
		time delays may simply be too long to measure in practice.

	\section*{Acknowledgment}
		
		For the inversions performed in this article we used the 
		infrastructure of the VSC - Flemish Supercomputer Center, funded by the 
		Hercules foundation and the Flemish Government - department EWI.


	\bsp 
	\label{lastpage}

\end{document}